\documentclass[twocolumn]{aastex631}

\shortauthors{A.S.Rajamuthukumar et al.}
\usepackage{multirow}
\usepackage{threeparttable}  
\usepackage{booktabs}
\usepackage{bm}
\graphicspath{{./}{figures/}}

\begin{document}

\title{Triple evolution: an important channel in the formation of type Ia supernovae}

\author[0000-0002-1872-0124]{Abinaya Swaruba Rajamuthukumar}
\affiliation{Max Planck Institut für Astrophysik, Karl-Schwarzschild-Straße 1, 85748 Garching bei München, Germany \\}

\author[0000-0003-1004-5635]{Adrian S. Hamers}
\affiliation{Max Planck Institut für Astrophysik, Karl-Schwarzschild-Straße 1, 85748 Garching bei München, Germany  \\}

\author[0000-0001-5853-6017]{Patrick Neunteufel}
\affiliation{Max Planck Institut für Astrophysik, Karl-Schwarzschild-Straße 1, 85748 Garching bei München, Germany  \\}

\author[0000-0003-3308-2420]{Rüdiger Pakmor}
\affiliation{Max Planck Institut für Astrophysik, Karl-Schwarzschild-Straße 1, 85748 Garching bei München, Germany  \\}

\author[0000-0001-9336-2825]{Selma E. de Mink}
\affiliation{Max Planck Institut für Astrophysik, Karl-Schwarzschild-Straße 1, 85748 Garching bei München, Germany  \\}
\affiliation{Anton Pannekoek Institute for Astronomy and GRAPPA, University of Amsterdam, NL-1090 GE Amsterdam, The Netherlands}



\begin{abstract}

 Type Ia supernovae (SNe Ia) are thought to be the result of thermonuclear explosions in white dwarfs (WDs). Commonly considered formation pathways include two merging WDs (the double degenerate channel), and a single WD accreting material from a H or He donor (the single degenerate channel). Since the predicted SNe Ia rates from WD in binaries are thought to be insufficient to explain the observed SNe Ia rate, it is important to study similar interactions in higher-order multiple star systems such as triple systems. We use the evolutionary population synthesis code Multiple Stellar Evolution (MSE) to study stellar evolution, binary interactions and gravitational dynamics of the triple-star systems. Also, unlike previous studies, prescriptions are included to simultaneously take into account the single and double degenerate channels, and we consider triples across the entire parameter space (including those with tight inner binaries). We explore the impact of typically ignored or uncertain physics such as fly-bys and CE prescription parameters on our results. The majority of systems undergo circular mergers to explode as SNe Ia, while eccentric collisions contribute to $0.4-4$ per cent of SNe Ia events. The time-integrated SNe Ia rate from the triple channel is found to be $(3.60 \pm 0.04) \times {10^{-4}}\,\mathrm{M_{\odot}}^{-1}$ which is, surprisingly, similar to that of the isolated binary channel where the SNe Ia rate is $(3.2 \pm 0.1) \times {10^{-4}}\,\mathrm{M_{\odot}}^{-1}$. This implies that triples, when considering their entire parameter space, yield an important contribution to the overall SNe Ia rate.

\end{abstract}



\section{Introduction} \label{sec:Introduction}

Type Ia supernovae (SNe Ia) are standard candles that play a key role in distance measurements on cosmological scales. As such, they play an important role in our understanding of the structure and expansion rate of the Universe. SNe Ia are important for our  comprehension of the chemical evolution of galaxies, and of iron-group element nucleosynthesis. The origin of SNe Ia is thought to be thermonuclear explosions in white dwarfs (WDs), though our understanding of their progenitors and explosion mechanisms is not very clear \citep{2012NewAR..56..122W,2014ARA&A..52..107M,2018PhR...736....1L,2020IAUS..357....1R}.

\begin{figure*}
    \centering
    \includegraphics[trim = 850 0 1050 0,width=1\textwidth]{figures/Intial_mass.pdf}
    \includegraphics[trim = 350 0 450 0,width=1\textwidth]{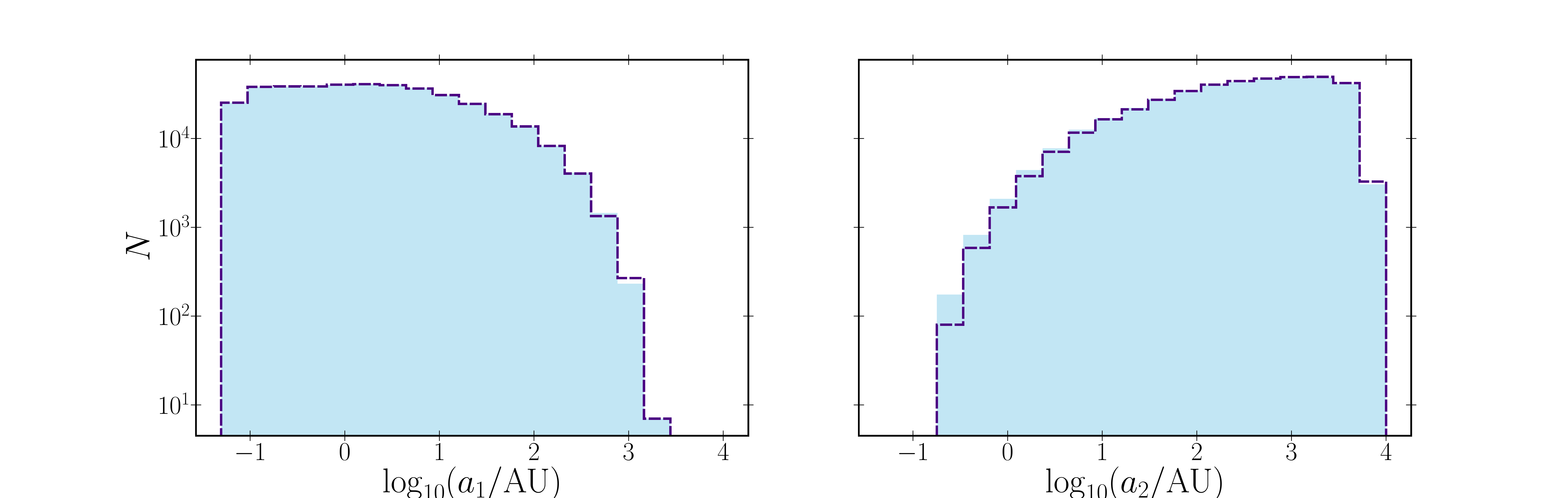}
    \includegraphics[trim = 350 0 450 0,width=1\textwidth]{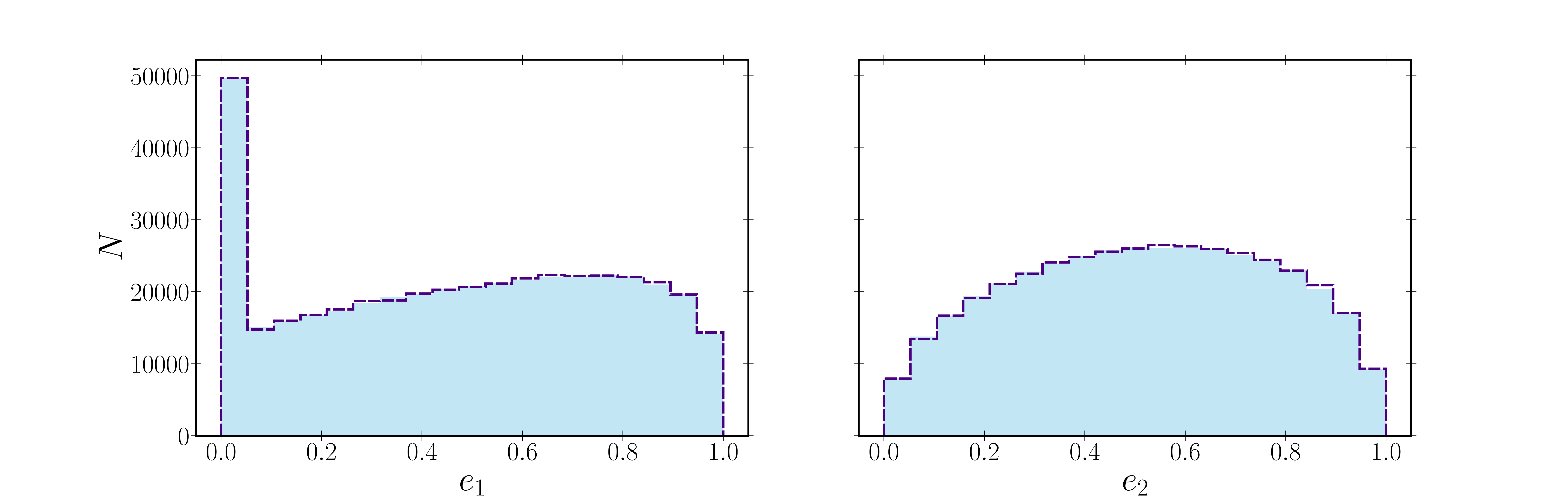}
    \caption{Initial parameter distributions of the constructed initial populations with two $q\mathrm{_{out}}$ models : the decaying exponential model (model 1, blue filled column) and the \citet{2017ApJS..230...15M} $q$ distribution (model 2, dashed indigo line). The top panels show the mass distributions of the primary, secondary, and tertiary star. The middle panels represent the semimajor axis distributions for the inner and outer orbits. The bottom panels show the eccentricity distributions for the inner and outer orbit.}
    \label{fig:img1}
\end{figure*}

Nevertheless, there are two commonly suggested binary progenitor channels  \citep{2020IAUS..357....1R} to produce SNe Ia explosions, which include the double degenerate (DD) channel (in which both the components of the binary system are WDs, \citealt{1984ApJ...277..355W}, \citealt{1984ApJS...54..335I}) and the single degenerate (SD) channel (in which an accreting Chandrasekhar mass WD may explode via delayed detonation, or an accreting sub-Chandrasekhar mass WD may explode via double detonation, \citealt{1973ApJ...186.1007W}). Although substantial amount of work \citep{1980SSRv...27..563N,1982ApJ...257..780N,1982ApJ...253..798N,2004A&A...419..623Y,2004A&A...419..645Y,2016A&A...589A..43N,2017A&A...602A..55N,2019A&A...627A..14N,2021ApJ...923L..34B} has been done to understand the progenitors through binary evolution, the general consensus is that the rates from the binary channels are too low to explain the observations \citep{2011MNRAS.417..408R,2014A&A...563A..83C}. The observed time integrated rate from \citet{2012MNRAS.426.3282M} is $(1.3 \pm 0.2) \times {10^{-3}}\,\mathrm{M_{\odot}}^{-1}$. \citet{2014A&A...563A..83C} studied the formation channels for SNe Ia through binary population synthesis and estimated the time integrated overall SN rate to be $4.5 \times {10^{-4}}\,\mathrm{M_{\odot}}^{-1}$, which could explain only a fraction of the observed rates from \citet{2012MNRAS.426.3282M}. This motivates the study of other progenitor channels. 

A formation channel for SNe Ia that has not been studied as extensively as the binary channel is the triple channel which involves hierarchical triple-star systems containing WDs. This channel is particularly interesting because of its contribution to producing and disrupting close binaries. For an isolated binary, it is difficult to produce close binaries and mergers within a Hubble time. However, in a triple system, if the initial mutual inclination is sufficiently large, the inner binary can undergo high-amplitude eccentricity oscillations. This in turn also leads to changes in the mutual inclination of the system. These oscillations are known as von Zeipel Lidov Kozai (ZLK) oscillations 
(\citealt{1910AN....183..345V,1962AJ.....67..591K,1962P&SS....9..719L}, see \citealt{2016ARA&A..54..441N} for a review). 
ZLK oscillations, combined with tidal effects, can shrink the inner binary which results in the formation of close binaries \citep{1979A&A....77..145M,2001ApJ...562.1012E,2006Ap&SS.304...75E,2007ApJ...669.1298F}, cause earlier CE evolution \citep{2019ApJ...882...24H,2020A&A...640A..16T}, accelerate mergers \citep{2002ApJ...578..775B,2011ApJ...741...82T,2018A&A...610A..22T}, and induce dynamical instability which in turn results in a merger of two WDs, and a SNe Ia explosion. From studies by \citet{2017ApJS..230...15M} and \citet{2010ApJS..190....1R}, we know that about $10\%$ of solar mass stars are found to be in triples with possible SNe Ia progenitors.

Previous studies of triple-star systems considered SNe Ia rates from head-on collisions arising from dynamical interactions \citep{2012arXiv1211.4584K}, contributions from WD mergers taking into account stellar evolution and dynamics \citep{2013MNRAS.430.2262H}, isolated triples with a circular approximation for mass transfer \citep{2018A&A...610A..22T}, the postulated progenitor triples from Gaia DR2 database \citep{2019MNRAS.490..657H}, and ultra-wide WD triples affected by fly-bys \citep{2019ApJ...882...24H, 2021MNRAS.500.5543M}. These studies considered dynamical interactions and/or stellar evolution and binary interactions such as tidal effects. However, their focus was on the DD formation pathway; there are no studies on the contribution of single degenerate channels (in which an accreting Chandrasekhar mass WD may explode via delayed detonation, or an accreting sub-Chandrasekhar mass WD may explode via double detonation, \citealt{1973ApJ...186.1007W}) in triple systems, and in particular also self-consistently taking into account stellar and binary evolution (especially mass transfer in eccentric orbits), as well as gravitational dynamics. 

In this paper, we present a comprehensive study of candidates for thermonuclear explosions originating from triple-star systems through both the single and DD channels. We note that not all thermonuclear SNe will result in SNe Ia, but may form related transients such as SNe Iax instead. For the purposes of this study, we use SN Ia as a catch-all form for transients resulting from the thermonuclear detonation of a WD. In addition, in our simulations, we take into account the possibility for the tertiary star to transfer mass onto to the inner binary system which in turn can produce a triple common envelope (TCE). The paper begins with the methodology in Section 2, followed by the different formation channels for SNe Ia in Section 3. In Section 4, we present our statistical results. We discuss and conclude the results in Sections 5 and 6, respectively.

\section{Methodology} \label{sec:Methodology}

\begin{figure*}
    \centering
    \includegraphics[trim = 150 100 120 0,  width=1\textwidth]{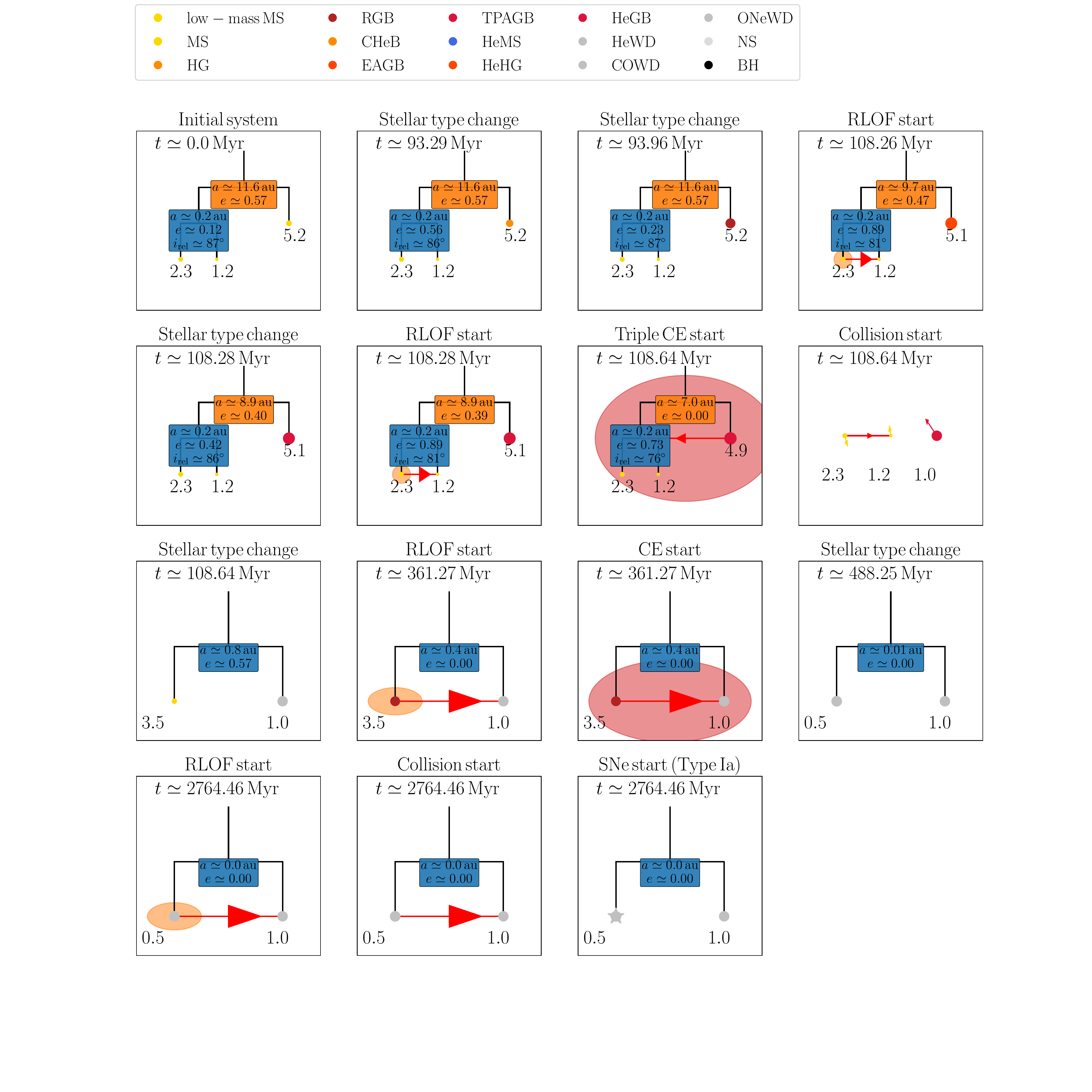}
    \caption{Example of a system undergoing triple common envelope (TCE). A triple system with a massive tertiary evolves first to transfer mass on top of the inner binary, to form a TCE. The inner binary merges to form a rejuvenated star at the end of TCE, which then interacts with the tertiary star to explode as SNe Ia. Refer to Appendix~\ref{appendix:mobilediagrams} for more details on the mobile diagrams presented in this paper.}
    \label{fig:img2}
\end{figure*}


\subsection{Population sysnthesis} \label{sec:Population synthesis}
\subsubsection{MSE} \label{sec:MSE}

In this work, we use the evolutionary population synthesis code MSE (Multiple Stellar Evolution; \citealt{2021MNRAS.502.4479H}, version-v0.87). The advantage of this code is that it incorporates prescriptions for stellar evolution, binary interactions (tides, mass transfer, etc.), dynamical perturbations from higher-order multiple systems, and fly-bys. MSE is a publicly available C/C++ code with a Python interface. It can evolve any number of stars as long as the system is originally hierarchical (later potential dynamical instabilities are modelled self-consistently through $N$-body methods). In order to tackle the complicated long-term dynamical evolution of multiple-star systems, MSE uses a hybrid approach which switches between the secular approximation \citep{2016MNRAS.459.2827H,2018MNRAS.476.4139H,2020MNRAS.494.5492H} and $N$-body integration \citep{2020MNRAS.492.4131R} during run time. Throughout the dynamical evolution, post-Newtonian (PN) terms are taken into account up to and including 2.5 PN order. 

Single star evolution in MSE is based on the SSE algorithms \citep{2000MNRAS.315..543H} based on stellar evolutionary tracks by \citet{1998MNRAS.298..525P}. The code uses modified BSE prescriptions \citep{2002MNRAS.329..897H} for binary interactions. Tidal evolution is modelled following the equilibrium tide model \citep{1998ApJ...499..853E}. Here the tides are applied to star-star and star-composite systems. The code takes into account eccentric mass transfer, adopting the model of \citet{2019ApJ...872..119H}. When mass transfer is deemed unstable, CE is modelled using the energy conservation mechanism, i.e, the $\alpha$-$\lambda$ CE prescription \citep{1976IAUS...73...75P}. The outer companion, when massive enough, can transfer mass onto the inner binary components, and the subsequent evolution is modelled following approximate prescriptions \citep{2022ApJS..259...25H}.

In MSE, the effects of passing stars (fly-bys) are taken into account as appropriate for low-density ($n_\star = 0.1\,\mathrm{pc}^{-3}$) environments. An exploration of the impact on triples of encounters in high-density environments such as globular clusters, although interesting, is beyond the scope of this paper. The perturber mass is sampled either from \citet{2001MNRAS.322..231K} and encounters are sampled assumed an encounter sphere of radius $R_\mathrm{enc} = {10^5}$ au with velocities sampled from Maxwellian distribution of dispersion $\sigma_\star = 30 \,\mathrm{kms}^{-1}$. These fly-bys become significant when the semimajor axis of the orbit exceeds approximately $10^3 \mathrm{\,au}$.

\begin{figure*}
    \centering
    \includegraphics[trim = 150 100 120 0 ,width=1\textwidth]{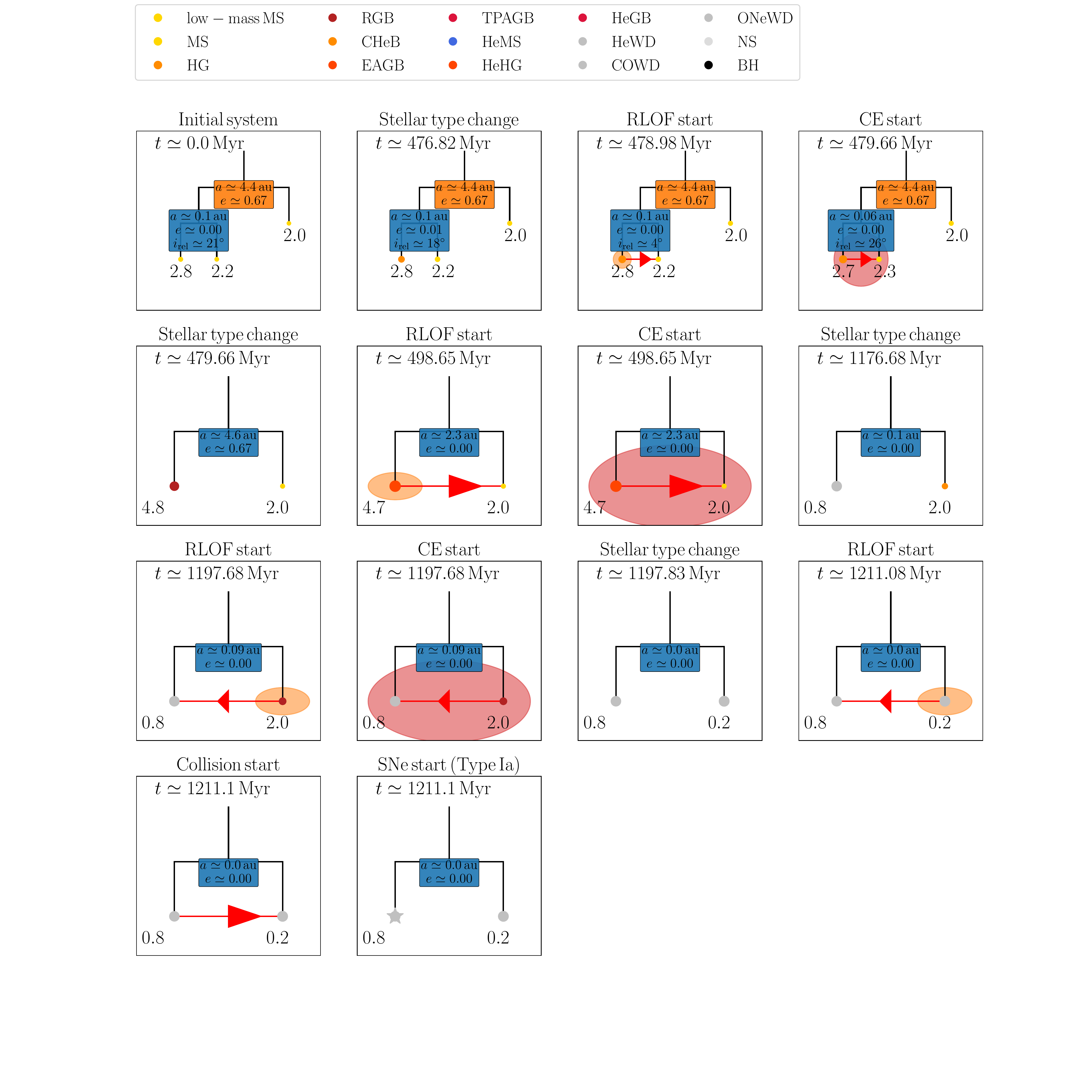}
    \caption{Example of a SNe Ia from the binary system that is formed as a result of double merger. A triple system in which the inner binary components merges to form a new massive star which then interacts with the tertiary star to produce SNe Ia explosion.}
    \label{fig:img3}
\end{figure*}

\subsubsection{SNe Ia prescription} \label{sec:SNe_pres}

The initial version of MSE \citep{2021MNRAS.502.4479H} uses BSE \citep{2002MNRAS.329..897H} prescriptions for SNe Ia explosions.  Within these prescriptions, an accreting CO WD has accumulated 0.15 $\mathrm{M}_\odot$ of of helium, the WD explodes in a SNe Ia. 
This assumption, however, has been shown to be incomplete since its first implementation. The amount of material required to initiate a helium detonation, and subsequent ignition of the CO-core has been shown to depend on other parameters of the progenitor binary, most notably the mass of the accretor, the mass transfer rate and, to some extent, assumptions on rotation, angular momentum transport and viscose heating \citep{2004A&A...419..645Y,2004A&A...419..623Y,2005A&A...435..967Y,2011ApJ...734...38W,2014MNRAS.445.3239P,2017A&A...602A..55N}. Further, as summarized particularly by \citet{2014MNRAS.445.3239P}, depending on the mass transfer rate, outcomes of He-accretion onto CO-WDs range from possible double detonation ($\dot{M} \lesssim 7\cdot10^{-8} \,\mathrm{M_\odot}/\mathrm{yr}$) via massive He-novae of decreasing intensity $\dot{M} \gtrsim 7\cdot10^{-8} \,\mathrm{M_\odot}/\mathrm{yr}$), \citep[see][]{2004ApJ...613L.129K} to steady burning and re ignition as a He-red giant ($\dot{M} \gtrsim 1 \cdot10^{-6} \,\mathrm{M_\odot}/\mathrm{yr}$) in a space of about two orders of magnitude. As further shown by \citet{2016A&A...589A..43N}, a system may move between these different mass transfer regimes, with systems, e.g., first undergoing weak helium flashes to then finally terminate in a SN. 
\begin{figure*}
    \centering
    \includegraphics[trim = 150 350 120 0 ,width=1\textwidth]{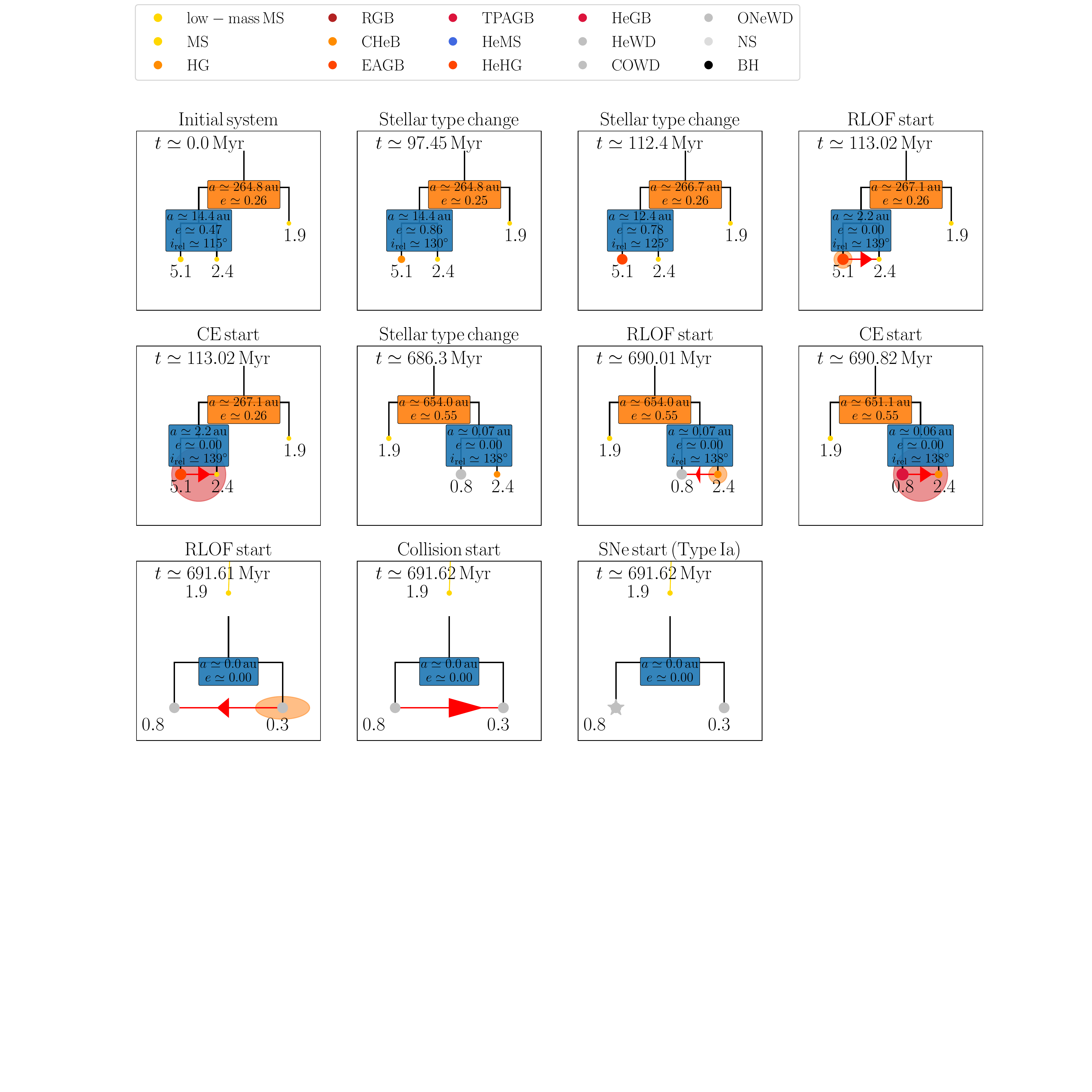}
    \caption{Example of an triple evolution channel in which the tertiary is unbound at the time of SNe Ia explosion. A triple system in which the primary star of the inner binary fills its Roche lobe and undergoes a first CE phase at around 113 Myr. At the end of CE, the primary evolves into a WD which then accretes mass from the secondary star forming a second CE. At the end of the second CE, the tertiary star has become unbound due to rapid mass loss in the inner orbit the secondary also evolves into a WD which then collides with the other WD to explode as SNe Ia.}
    \label{fig:img4}
\end{figure*}

In order to take these different possibilities into account, this study employs a refined prescription, considering the mass of the accretor as well as the rate of mass accretion, for deciding on the final outcome of helium accretion \citet{2016A&A...589A..43N}. This prescription combines the accretion-rate-dependent accretion efficiencies ($\eta$) presented by \citet{2004ApJ...613L.129K} at values of mass transfer rates with the occurrence of detonation at low mass transfer rates as presented by \citet{2011ApJ...734...38W}. The resulting prescription can be written
\begin{equation} \label{eq:acceff_prescription}
  \eta = \left\{ \begin{array}{ll}
     1, & \mathrm{if}~ 0 < [\dot{M}] < \dot{M}_{\mathrm{WK,max}}~,\\
     0, & \mathrm{if}~ \dot{M}_\mathrm{WK,max} < \dot{M} <  \dot{M}_{\mathrm{KH,min}}~,\\
    \eta_\mathrm{KH}(M_{\mathrm{WD}},\dot{M}), &  \mathrm{if}~ \dot{M}_{\mathrm{KH,min}} < \dot{M}~,\\
    1, & \mathrm{if}~ \dot{M}_{\mathrm{KH,max}} < \dot{M}
  \end{array} \right.
\end{equation}
with $\dot{M}_{\text{WK,max}}$ and $\dot{M}_{\text{KH,min}}$ the upper and lower limits of accretion rates studied by \citet{2011ApJ...734...38W} and \citet{2004ApJ...613L.129K} respectively. We note that, while \citet{2016A&A...589A..43N} used a time-averaged mass transfer rate in order to exclude ignitions resulting from spurious variations in the mass transfer rate, as are prone to happen in detailed stellar evolution, this approach is unnecessary in the context of population synthesis.

With regards to the DD SNe Ia, a new prescription combining results from varies previous hydrodynamics simulation studies has been incorporated into MSE. Collisions in MSE can happen either via a circular merger, usually following CE evolution, or eccentric collision driven by secular evolution. We assume the outcome will be SNe Ia when there is a circular merger of a He WD and CO WD (irrespective of their masses). We also assume that the coalescence of two CO WDs in which one of them is more massive than 0.9 $\mathrm{M}_\odot$ results in a SNe Ia explosion \citep{2010Natur.463...61P,2013ApJ...770L...8P}. In the event of an eccentric collision, the collision of two CO WDs, two ONe WDs, or a CO WD and an ONe WD is assumed to lead to SNe Ia.

\subsection{Initial distributions\label{subsec:Initial distributions}}
\begin{figure*}
    \centering
    \includegraphics[trim = 150 350 120 0 ,width=1\textwidth]{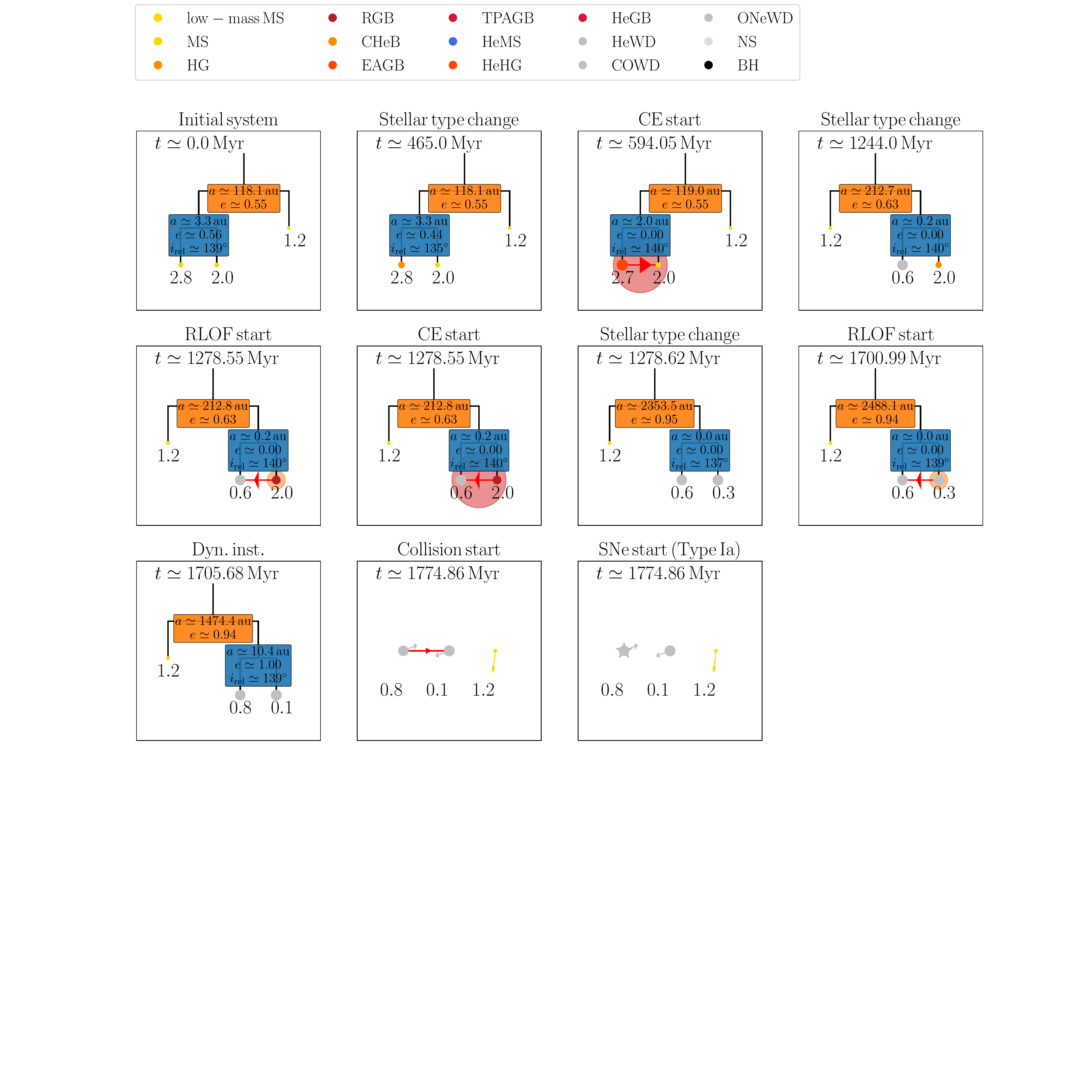}
    \caption{Example of an eccentric collision. The isolated binary channel usually predicts that the binary gets circularized ($e \approx 0$) after a CE episode. Here, the figure shows that the inner binary can still achieve high eccentricities from dynamical perturbations, which then can lead to a SNe Ia explosion via collision. }
    \label{fig:img5}
\end{figure*}

We adopt a population synthesis method, in which the initial conditions for a large number of triple-star systems are generated based on a Monte Carlo approach. Here, we describe the assumptions made in this procedure. The primary mass of the inner binary $m_1$ (i.e., the mass of the initially most massive star in the inner binary system), is set between 1 $\mathrm{M}_\odot$ and 6.5 $\mathrm{M}_\odot$ to ensure the formation of at least one CO WD within a Hubble time in isolation, and it follows \citet{2001MNRAS.322..231K}. The distribution of the secondary mass is modeled after the observational fit functions of \citet{2017ApJS..230...15M}. The initial orbital period and eccentricity distributions (for both inner and outer orbits) are also drawn from the observational fit functions of \citet{2017ApJS..230...15M}. The orbital periods are sampled in the range $0.2 < \mathrm{log}  (P/\mathrm{days}) < 8$. Eccentricities of both orbits are sampled between 0 and 1. The initial mutual inclinations are uniformly distributed in $\cos(i)$. The longitudes of the ascending node and arguments of periapsis  are sampled from uniform distributions. These assumptions correspond to isotropic orientations of the inner and outer orbits. The systems that does not satisfy the stability criteria of \citet{2001MNRAS.321..398M} are rejected. We also eliminate systems with stars that are filling their Roche lobes at the start of the evolution at periapsis, using the fit of \citet{1983ApJ...268..368E}, and using the mass-radius relation $R  \propto M^{0.7}$ to estimate the initial stellar main-sequence radii\footnote{This more approximate method of determining the main-sequence radii is only adopted for sampling purposes.}.

\begin{table*}
\caption{Overview of the different models, stating the assumptions for the distribution of the mass ratio between the outer star and the inner binary $q_\mathrm{out} \equiv m_3/(m_1+m_2)$, the choice for the CE paramater $\alpha_{ce}$, whether or not fly-bys are accounted for, and the metallicity}
\begin{center}
\begin{tabular}{c c c c c} 
\toprule
Models & $q_{\mathrm{out}}$ & $\alpha_{\mathrm{CE}}$ & Fly-bys & Metallicity \\
\midrule
Model 1 &  $\exp(-q_\mathrm{out}\lambda)$; $\lambda = 1.05$ &  1 & Included & 0.02 \\
Model 2 & Extrapolating \citealt{2017ApJS..230...15M}  & 1  & Included & 0.02 \\
Model 3 & $\exp(-q_\mathrm{out}\lambda)$; $\lambda = 1.05$ & 10 & Included & 0.02 \\
Model 4 & $\exp(-q_\mathrm{out}\lambda)$; $\lambda = 1.05$ & 0.1 & Included & 0.02 \\
Model 5 & $\exp(-q_\mathrm{out}\lambda)$; $\lambda = 1.05$ & 1 & Ignored & 0.02 \\
Model 6 & $\exp(-q_\mathrm{out}\lambda)$; $\lambda = 1.05$ & 1 & Included & 0.001 \\
\bottomrule
\end{tabular}
\end{center}
\label{table:Table 1}
\end{table*}

\begin{figure*}
    \centering
    \includegraphics[trim = 150 600 120 0 ,width=1\textwidth]{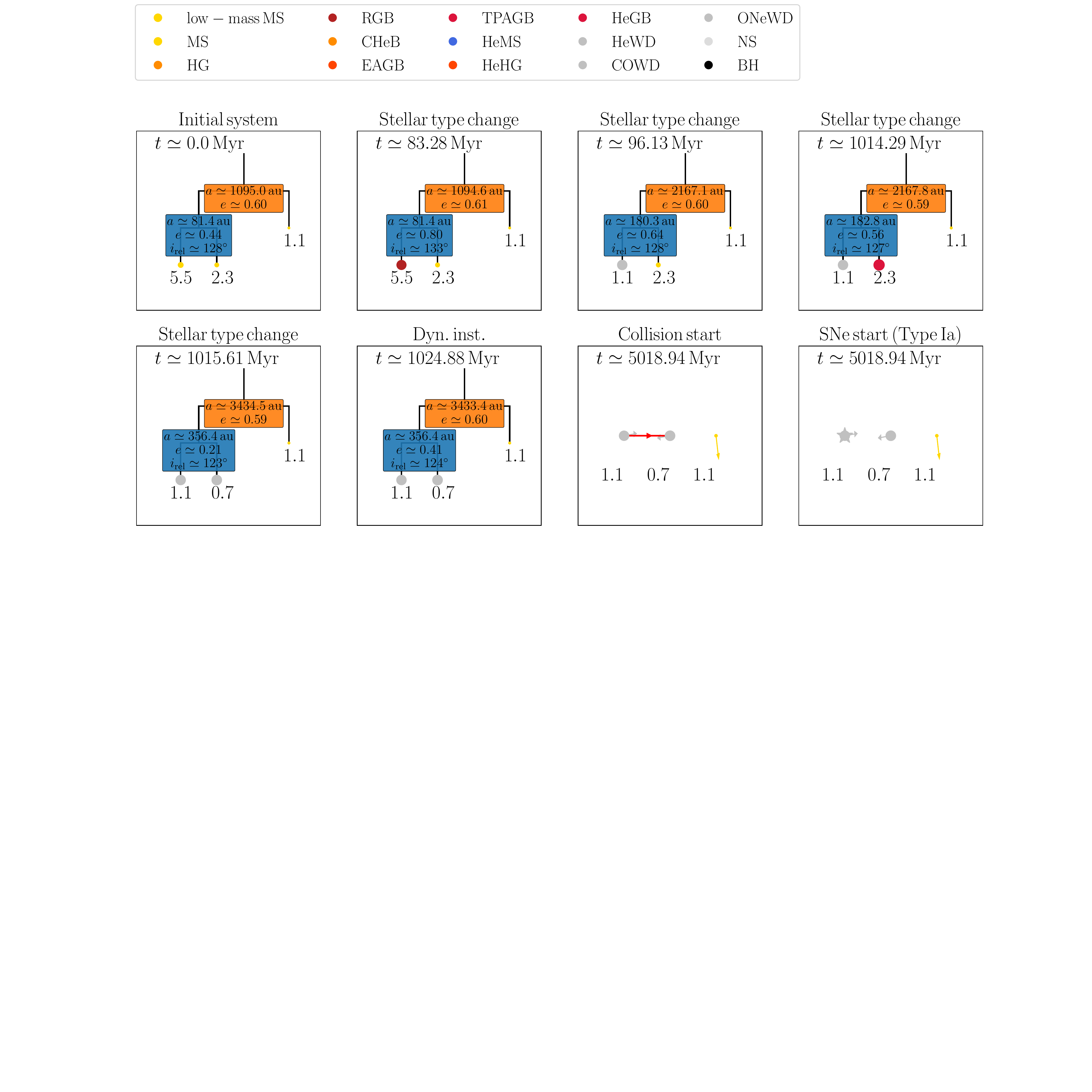}
    \caption{Example of a SNe Ia driven by dynamical instability, without any CE evolution phases. A triple system which undergoes eccentricity oscillations in the inner binary, then the inner binary WDs collide due to dynamical instability, producing a SNe Ia explosion. This channel is particularly interesting in the sense that it does not involve any CE phase.}
    \label{fig:img6}
\end{figure*}

 The mass-ratio distribution involving the tertiary (outer) star in triple systems, specifically, the outer mass ratio $q_\mathrm{out} \equiv m_3/(m_1+m_2)$, is not very well constrained. From the Multiple Star Catalogue (MSC; \citealt{2018ApJS..235....6T}), there are about seven per cent of systems in which the tertiary star is more massive than the total mass of the inner binary. These systems are potentially interesting because they favor channels which involve triple mass transfer, i.e., when the tertiary star fills its Roche lobe around the inner binary \citep{2021MNRAS.500.1921G,2022ApJS..259...25H}. In order to allow for the possibility of systems with a massive tertiary star, we have constructed two different models for the initial outer mass ratio distribution. For the first model (hereafter Model 1), we fit a decaying exponential function to the data from the MSC and find the best fitted parameters. The second model (hereafter Model 2) is an extrapolation of the mass ratio distribution of \citet{2017ApJS..230...15M}.  The first mass ratio model is of the form

 \begin{equation}
    \frac{\mathrm{d}N}{\mathrm{d}q_\mathrm{out}} \propto \exp(-q_\mathrm{out}\lambda), 
\end{equation}
 where $\lambda$ = 1.05. The decaying exponential model best fits the current observations. However, the MSC has substantial observational biases for triple (and higher-order) systems and we especially expect strong observational biases against triples with high-mass tertiaries and low mass inner binaries (high mass-ratio systems). Thus, we take into account both models in our work, as a means to explore the current uncertainties in the outer mass ration distribution.
 
 In addition to considering two models for the assumed distribution of $q_\mathrm{out}$, we vary physical model parameters in our simulations to investigate the impact of physical uncertainties, as well as effects that are often ignored in the literature. We inspect the effects of fly-bys and various CE parameters in our work. We also study the SNe Ia rate from stars with sub-solar metallicity.
 
 \subsection{Construction of the initial population} \label{sec:pop_construction}

Our population pool includes $4 \times 10^5$ triple systems for both Models 1 and 2. Triple systems which include all the varying model parameters such as CE, flybys and metallicity constitute $4 \times 10^5$ systems. In total, our triple population sample size sums up to be $1.2 \times 10^6$. In addition, in order to investigate the effect of the tertiary star, we re-run our main models without the tertiary star (only inner binary systems). The size of the inner binary population is $8 \times 10^5$. We also study the contribution from isolated binaries, for which we construct a binary population of size $1 \times 10^5$. As explained later in Section 4.4, the latter isolated binary population is significantly different from the triple population with the tertiary star removed. In total, our population pool consists of $2.1 \times 10^6$ systems. The constructed population is evolved for a period of 10 Gyr with an imposed maximum wall time of 5 hours.

\section{Evolutionary Pathways} \label{subsec:Evolutionary Pathways}
In this section, we summarize evolutionary pathways for forming SNe Ia in triple systems as found in our population synthesis calculations. We restrict our explanation to formation channels that demand a tertiary to form SNe Ia. In order to select the systems that have a effect from the tertiary, we compare SNe Ia from triple population synthesis with those from inner binary (without tertiary) population synthesis. We provide 5 unique formation channels for producing SNe Ia only from triple systems. Table\,\ref{table:Table 2} quantifies the contribution from these evolutionary pathways. The presented evolutionary pathways are unique to triples and not mutually exclusive. The evolutionary pathways in which the tertiary does not contribute in producing SNe Ia explosion is similar to binary evolution channels and not being presented here.

\subsection{Triple common envelope\label{subsec:TCE}}
From our results, TCE is an important channel for producing SNe Ia from triple-star systems, responsible for $4-13\%$ of all SNe Ias by triples in our set of models. Fig.\,\ref{fig:img2} shows a mobile diagram (see Appendix\,\ref{appendix:mobilediagrams} for an explanation of the mobile diagrams presented here) of a triple system undergoing TCE, causing a merger of the inner binary, which then leads to a SNe Ia later. 

If the tertiary star is relatively close and more massive than the total mass of the inner binary, it can start transferring mass on to the inner binary, forming a TCE around the inner binary. At the end of TCE, if dynamical instability is triggered, one of the inner binary components can get exchanged with the tertiary, forming an exchange triple. Other possibilities include TCE evolution followed by a merger of the inner binary, or a merger of an inner binary component with the tertiary. TCE evolution can disrupt the triple system by unbinding the tertiary or inner binary component, resulting in a binary system with the remaining components. If the CE is assumed to be more efficient ($\alpha_\mathrm{CE} = 10$), it induces more inner binary mergers and thereby fewer TCE episodes.

\subsection{Double mergers\label{subsec:Double mergers}}
We identify two different cases of scenarios leading to SNe Ia and involving double mergers. In the first case, there is an early mass-transfer episode during the main sequence, which merges the inner binary components to a rejuvenated main-sequence star. This results in a new binary, with one the component being the original tertiary star, and the other as the merger remnant of the inner binary. These two stars then evolve, and this later leads to a SNe Ia explosion. In the second case, the tertiary star, as a result of secular eccentricity excitation coupled with tides, shrinks the inner binary which leads to an early CE phase that subsequently merges the inner binary. This forms a new binary with the merger remnant and original tertiary star, and produces a SNe Ia event later. Fig.\,\ref{fig:img3} shows a triple-star system in which the inner binary merges through a CE phase to produce a new star, which then further undergoes two more CE phases and then collide with the tertiary star to produce SNe Ia.

\begin{figure*}
    \centering
    \includegraphics[width=1\textwidth]{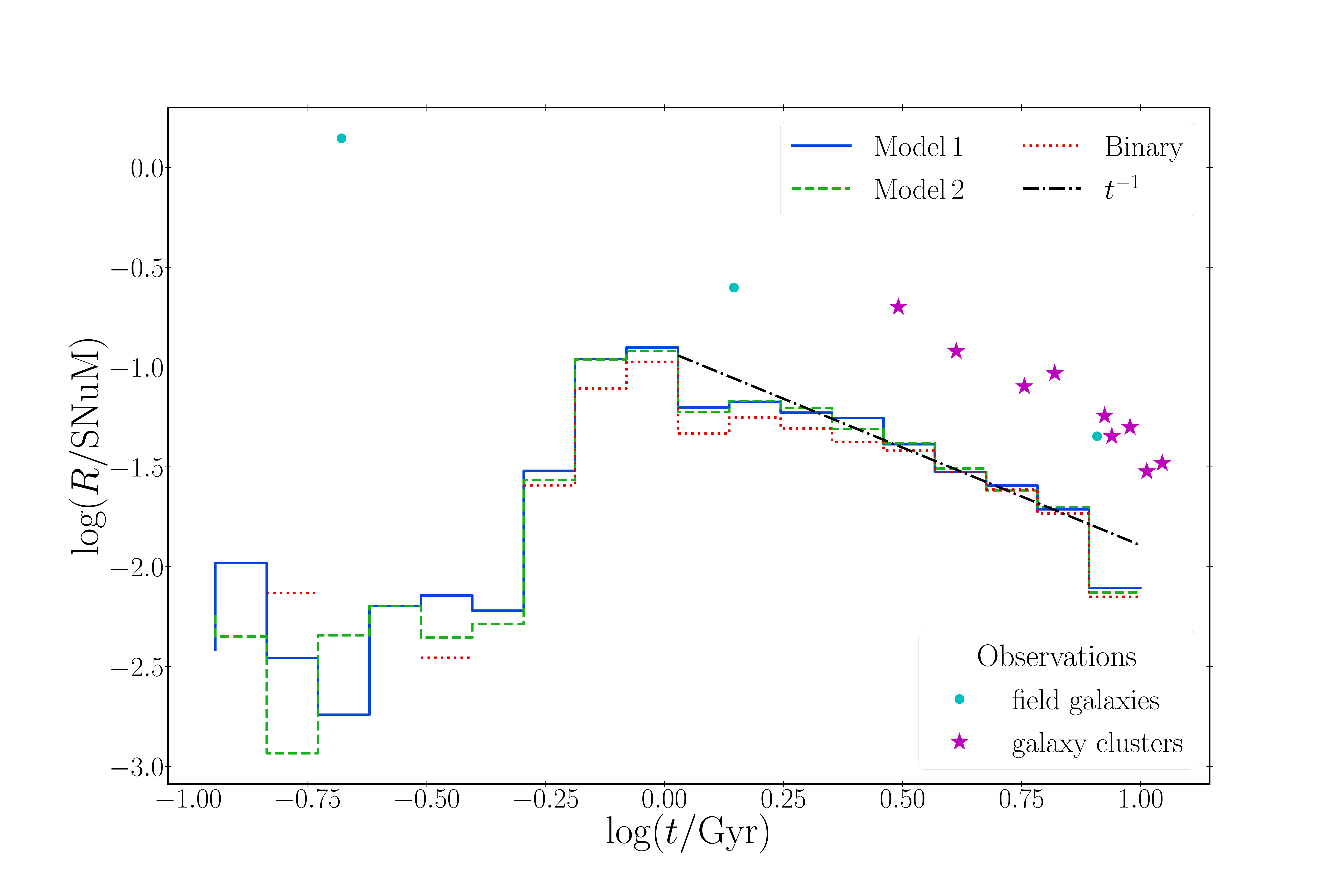}
    \caption{Delay time distribution (DTD) from all DD SNe Ia in our simulations. The solid blue line (Decaying exponential $q\mathrm{_{out}}$ model) and dashed green line (Extrapolating $q$ from \citet{2017ApJS..230...15M} $q\mathrm{_{out}}$ model) correspond to the DTD from triple population synthesis. The dotted red line represents the DTD from isolated binary population synthesis. Also, the black dashed dot line shows that the later parts of the DTD is found to follow a power law shape ${(\propto {t^ {-1}})}$. In addition, observational SNe Ia rates from field galaxies (Refer to Table 2 from \citet{2017ApJ...848...25M}) and galaxy clusters (Refer to Table 3 from \citet{2017ApJ...848...25M}) are shown by cyan and magenta points, respectively.}
    \label{fig:img7}
\end{figure*}

\subsection{Unbound tertiary\label{subsec: unbound tertiary}}
In the course of the evolution, the tertiary star can get unbound due to different reasons. For example, Fig.\,\ref{fig:img4} shows a triple channel in which the tertiary gets unbound when the inner binary undergoes a CE phase. CE in inner binary  and TCE are responsible for unbinding the tertiary star in about 67 per cent and 7 per cent of SNe Ia events through this channel, respectively. When the tertiary star is massive enough, it can collapse into a neutron star resulting in a type II supernovae. In about 22 per cent of the Unbound tertiary channel, the mass loss and/or natal kick during this type II supernovae can unbind the tertiary star, while the inner binary later produces a SNe Ia event. The tertiary star can also become unbound when there is a CE episode in the inner binary, which is associated with rapid mass loss in the inner binary. Fly-bys unbind the tertiary star when the semimajor axis is of the order of $10^3$ au or wider. When the triple system becomes dynamically unstable, one of the stars can get ejected out of the system. Fly-bys and dynamical unstability contribute 2 per cent each to this channel.

\subsection{Eccentric collision\label{subsec:Eccentric collision}}
The formation of close binaries in the isolated binary channel is mainly explained by CE phases. These systems are nearly always circularized at the end of the CE phase. But in the case of triple star systems, there is a possibility that, even after the CE phase, the tertiary can induce eccentricities in the close inner binary through secular evolution. We see that about 1 per cent of systems that form SNe Ia, experience eccentric collisions. There is also another possibility of forming a SNe Ia only through the dynamical channel. In these type of systems, the eccentricity of the inner binary oscillate due to the perturbations from the tertiary star and after an elapse of time, the secular approximation breaks down \citep{2014ApJ...781...45A,2014MNRAS.439.1079A,2016MNRAS.458.3060L}, the inner binary components collide at extremely-high eccentricity, leading to a SNe Ia explosion. Fig.\,\ref{fig:img5} shows an example of a system achieving such high eccentricities $(e \approx 1$), then causing a collision and hence a SNe Ia explosion.

\subsection{SNe Ia driven by dynamical instability\label{subsec:dynamical_unstability}}
In an isolated binary channel, the formation of SNe Ia cannot be explained without undergoing a CE phase. In our simulations, in addition to mergers during CE, there is also the possibility of physical collisions driven by (chaotic) few-body dynamics, following the onset of dynamical instability. For example,  Fig.\,\ref{fig:img6} shows a pure dynamical channel to produce SNe Ia. During the course of the evolution, if the inner binary becomes dynamically unstable following \citet{2001MNRAS.321..398M}, there can be a head-on collision between the binary components which leads to a SNe Ia explosion.

\section{Statistical Results} \label{sec:statistical results}
We present the delay time distribution and detailed statistical analysis of the contribution from different progenitors in this section. Table\,\ref{table:Table 3} gives the overview of the time-integrated rate from triple (various models) and binary channels. Table\,\ref{table:Table 4} summarizes the contributions of different progenitors to SNe Ia events.

\subsection{Delay time distribution and SNe Ia rate\label{subsec:DTD}}

We assume a starburst at time $t=0$; the SNe Ia rate during a particular time interval $\Delta T$ is then calculated using 
\begin{equation}
    R  = \frac{N}{{M_\star \Delta T} },
\end{equation}
where $N$ is the total number of SNe Ia explosions during $\Delta T$ and $M_\star$ is the total mass of the synthesized stellar population. We assume that the synthesized stellar population only constitutes of single, binary and triple stars. And, any contribution from  higher order systems is neglected. We use the primary mass dependent multiplicity fraction from \citet{2017ApJS..230...15M} (Refer to Table 13 from \citealt{2017ApJS..230...15M}) from when calculating the total stellar mass. 

Firstly, the total number of systems to be sampled are calculated using the following expression

\begin{widetext}
\begin{equation}    
    {N}_\mathrm{tot}  ={N_\mathrm{triple}} + {N_\mathrm{binary}} + {N_\mathrm{single}}  = \frac{{N_\mathrm{calc}}}{{F_\mathrm{calc}}} +  \sum_{m \in m_\mathrm{bins}}{N_{\mathrm{triple},\,m}} \frac{{{\alpha_{\mathrm{binary},\,m}}}}{{\alpha_{\mathrm{triple},\,m}}} +  \sum_{m \in m_\mathrm{bins}}{N_{\mathrm{triple},\,m}} \frac{{{\alpha_{\mathrm{single},\,m}}}}{{\alpha_{\mathrm{triple},\,m}}} ,
\end{equation}
\end{widetext}
where ${F_\mathrm{calc}}$ is the numerically calculated fraction of triple stars in which the primary mass of the inner binary is in the mass range 1-6.5 $\mathrm{M}_\odot$ and ${N_\mathrm{calc}}$ is number of triple stars originally sampled (in our case, $4 \times 10^5$ for each ${q_\mathrm{out}}$ model). ${N_{\mathrm{triple},\, m}}$ is the number of triple stars in the particular mass bin with  ${{\alpha_{\mathrm{single},\,m}}}$, ${{\alpha_{\mathrm{binary},\,m}}}$, and  ${{\alpha_{\mathrm{triple},\,m}}}$ being the single, binary, and triple fractions in the respective mass bins which we adopt from \citet{2017ApJS..230...15M}. A stellar population is constructed with the calculated number of single, binary and triple stars in each mass bin. The single star population is created by assuming the IMF from the \citet{2001MNRAS.322..231K} between 0.08 $\mathrm{M_\odot}$ and 100 $\mathrm{M_\odot}$. The primary mass of the binary population is constructed similarly to the single star population. The separations are calculated as a function of primary mass following \citet{2017ApJS..230...15M}. The secondary mass and eccentricities are sampled following the primary mass and period-dependent distribution functions from \citet{2017ApJS..230...15M}. The triple population is constructed similarly as described in Section\, \ref{sec:pop_construction}, except that now the primary masses of the inner binary are sampled in the mass range 0.08-100 $\mathrm{M}_\odot$, to cover the entire mass range. Finally, $M_\star$ is calculated by adding all the stellar masses.

\begin{table*}
\begin{center}
\begin{tabular}{c c c c c c} 
\toprule
Models &  TCE & Double mergers & Unbound Tertiary  & Eccentric collision & Dynamical Instability\\
 &  $(\%)$ & $(\%)$ & $(\%)$ & $(\%)$ & $(\%)$ \\
\midrule
{Model\,1}&$12.0 \pm 0.5$& $21.4 \pm 0.5$ & $28.1 \pm 0.7$ & $0.8 \pm 0.1$ & $1.1 \pm 0.1$\\
{Model\,2}&$12.6 \pm 0.5$& $20.1 \pm 0.5$ & $29.9 \pm 0.8$ & $1.1 \pm 0.1$ & $1.0 \pm 0.1$\\
{Model\,3}& $3.8 \pm 0.6$ & $20.1 \pm 1.0$ & $31.2 \pm 1.9$ & $4.0 \pm 0.6$ & $2.3 \pm 0.5$\\
{Model\,4}& $12.6\pm 1.8$ & $60.0\pm 3.4$ & $14.3 \pm 1.9$ & $2.3 \pm 0.7$ & $2.1 \pm 0.7$\\
{Model\,5}& $11.8 \pm 0.9$ & $20.1 \pm 1.0$ & $26.5 \pm 1.4$ & $0.4 \pm 0.1$ & $0.7 \pm 0.2$\\
{Model\,6}& $11.7 \pm 0.9$& $21.0 \pm 1.0$ & $27.8 \pm 1.5$ & $0.9 \pm 0.2$ & $1.1 \pm 0.3$\\
\bottomrule
\end{tabular}
\end{center}
\caption{Relative contribution of the formation channels for SNe Ia in triples. The channels listed in the table are the ones in which the tertiary plays a role in producing SNe Ia events. Note that they are not mutually exclusive. Error bars indicate statistical (Poisson) uncertainties.}
\label{table:Table 2}
\end{table*}

\begin{table}
\begin{tabular}{c c  c} 
    \toprule
    Models & Channel & SNe Rate\\
     &  & ($10^{-4}\mathrm{M_\odot}^{-1}$) \\
    \midrule
    \multirow{2}{*}{Model\,1}&Triple & $3.60 \pm 0.04$ \\
    & Inner Binary & $2.90 \pm 0.04$  \\
    \hline
    \multirow{2}{*}{Model\,2}&Triple& $3.50 \pm 0.04$ \\
    &Inner Binary & $2.90 \pm 0.04$ \\
   \midrule
    {Model\,3}&Triple & $2.40 \pm 0.07$ \\
    {Model\,4}&Triple& $0.90 \pm 0.04$ \\
    {Model\,5}&Triple&$3.70 \pm 0.09$ \\
    {Model\,6}&Triple& $3.60 \pm 0.09$ \\
    \midrule
    {Isolated Binary}&& $3.2 \pm 0.1$ \\
    \midrule
    {Observed rate}&& $13.0 \pm 1.5$ \\
    \bottomrule
\end{tabular}
\caption{SNe rate from different channels (triple, inner-binaries of triples with the tertiary removed, and isolated binaries) and models.}
\label{table:Table 3}
\end{table}

\begin{figure}
 \includegraphics[width=1\columnwidth]{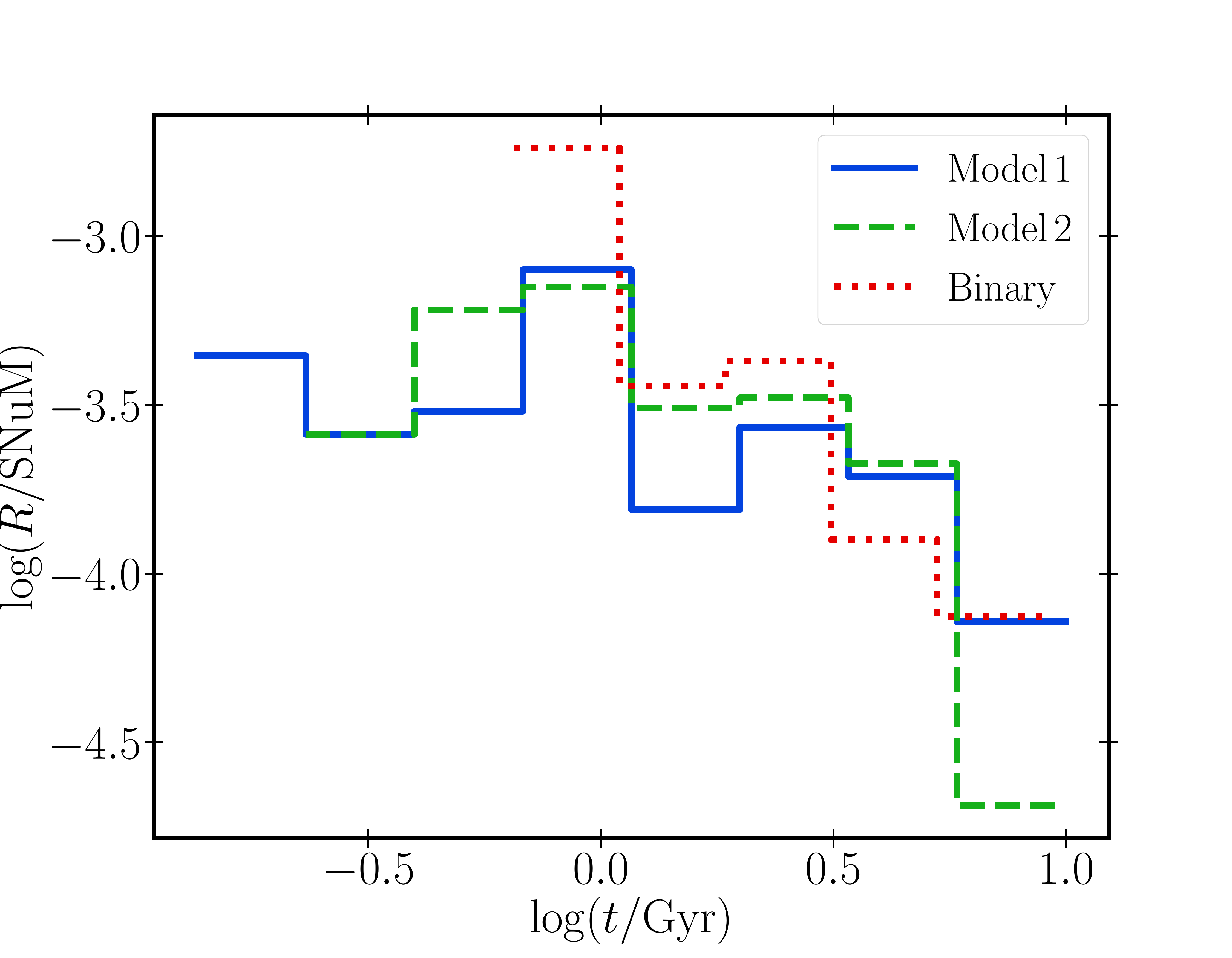}
 \caption{DTD from SD SNe Ia. The solid blue line ($q\mathrm{_{out}}$ model - Decaying exponential $q\mathrm{_{out}}$ model) and dashed green line ($q\mathrm{_{out}}$ model - Extrapolating $q$ from \citet{2017ApJS..230...15M}) correspond to the DTD from our simulated triples. The dotted red line represents the DTD from our simulated isolated binaries.}
 \label{fig:img8}
\end{figure}

The delay time distribution (DTD) for the DD and SD pathways are shown in Fig.\,\ref{fig:img7} and Fig.\,\ref{fig:img8}, respectively. The later part of the DD DTD is found to follow a power law shape, $\propto t^{-1}$. The delay time distribution is calculated in the units of SNuM (number of SNe Ia per $\mathrm{10^{10}\,M_{\mathrm{\odot}}}$ per century).

The time integrated SNe Ia rate $N_\mathrm{total}/M_\star$  for ${\mathrm{Model\,1}}$ and ${\mathrm{Model\,2}}$ are calculated as $(3.60 \pm 0.04) \times {10^{-4}}$ $\mathrm{M}_\odot^{-1}$ and $(3.50 \pm 0.04) \times {10^{-4}}$ $\mathrm{M}_\odot^{-1}$, respectively.

\subsection{Circular and eccentric mergers\label{subsec:Cir_ecc}}

\begin{figure*}
    \centering
    \includegraphics[width=1\textwidth]{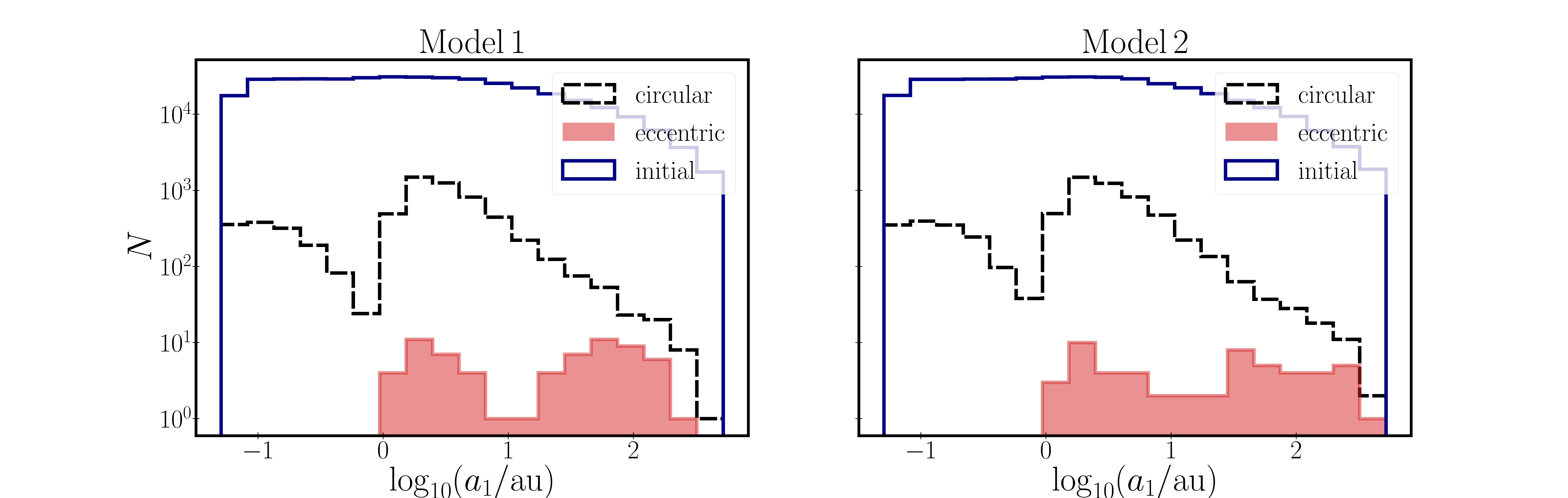}
    \caption{Distribution of the initial inner semimajor axes of all triple systems (blue solid lines),that explode as SNe Ia via circular mergers (black dashed lines) and eccentric collision (red filled columns). The two models represent the results from the two $q\mathrm{_{out}}$ models: Model 1 - Decaying Exponential fit to $q\mathrm{_{out}}$ observations, Model 2 - Extrapolating \citet{2017ApJS..230...15M} $q$ distribution.}
    \label{fig:img9}
\end{figure*}

\begin{figure*}
    \centering
    \includegraphics[width=1\textwidth]{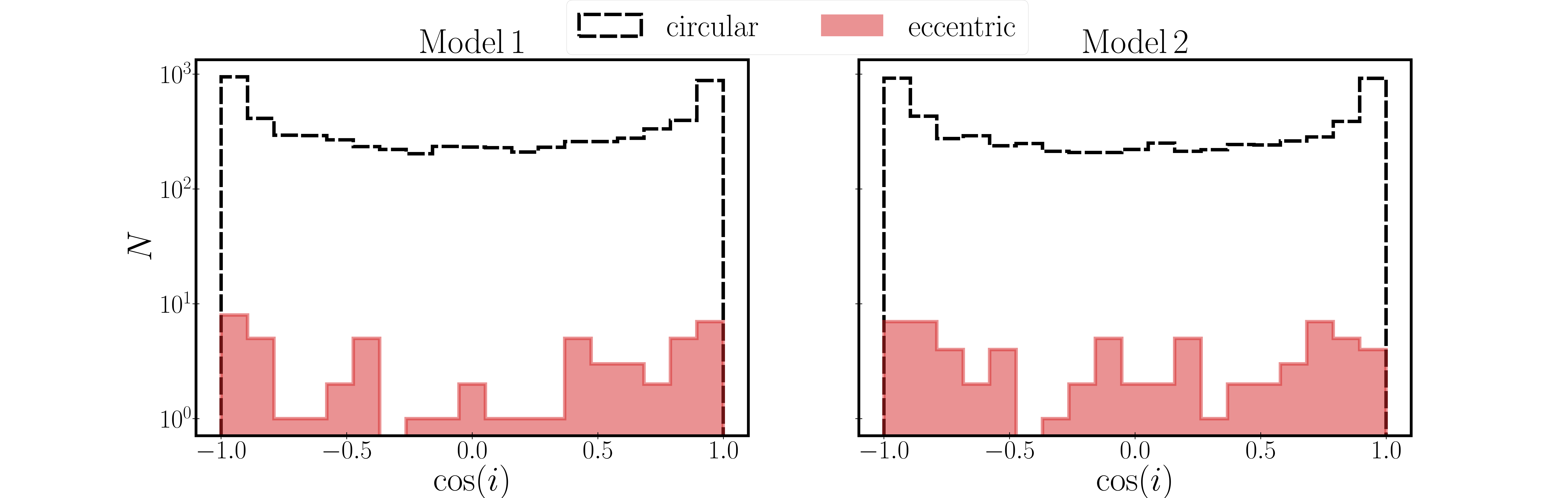}
    \caption{Distribution of the mutual inclinations of triple systems that explode as SNe Ia via circular mergers (black dashed lines) and eccentric collisions (red filled columns). The two models represent the results from the two $q\mathrm{_{out}}$ models: Model 1 - Decaying Exponential fit to $q\mathrm{_{out}}$ observations, Model 2 - Extrapolating \citet{2017ApJS..230...15M} $q$ distribution.}
    \label{fig:img10}
\end{figure*}

\begin{figure}
 \includegraphics[width=1\columnwidth]{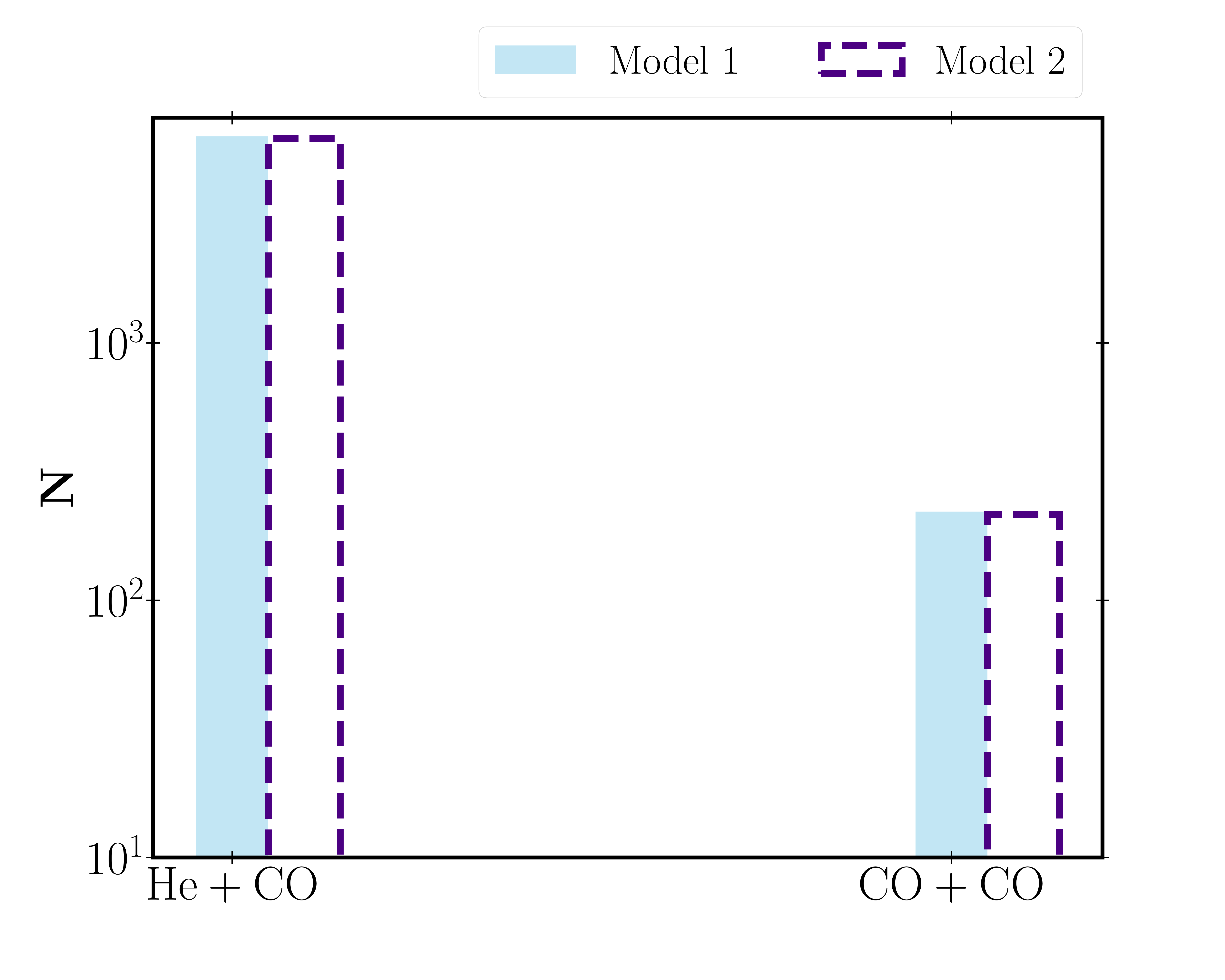}
 \caption{Combinations of mergers/collisions of He+CO WDs and CO+CO WDs contributing to DD SNe Ia.}
 \label{fig:img11}
\end{figure}

\begin{figure}
 \includegraphics[width=1\columnwidth]{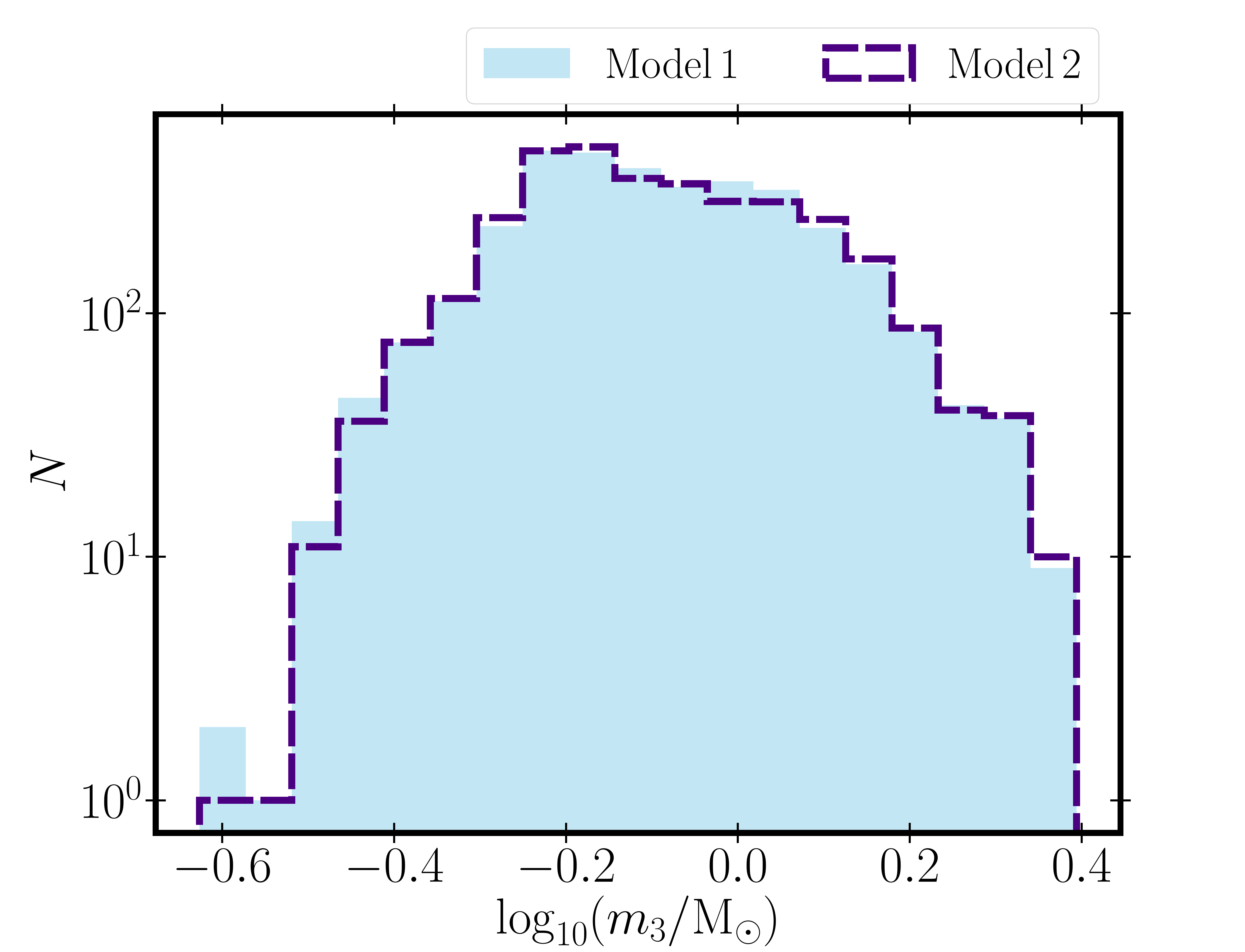}
 \caption{Distribution of the tertiary star mass when the inner binary explodes as SNe Ia, in those cases when the former is still bound at the time of the explosion.}
 \label{fig:img12}
\end{figure}

\begin{figure}
 \includegraphics[width=1\columnwidth]{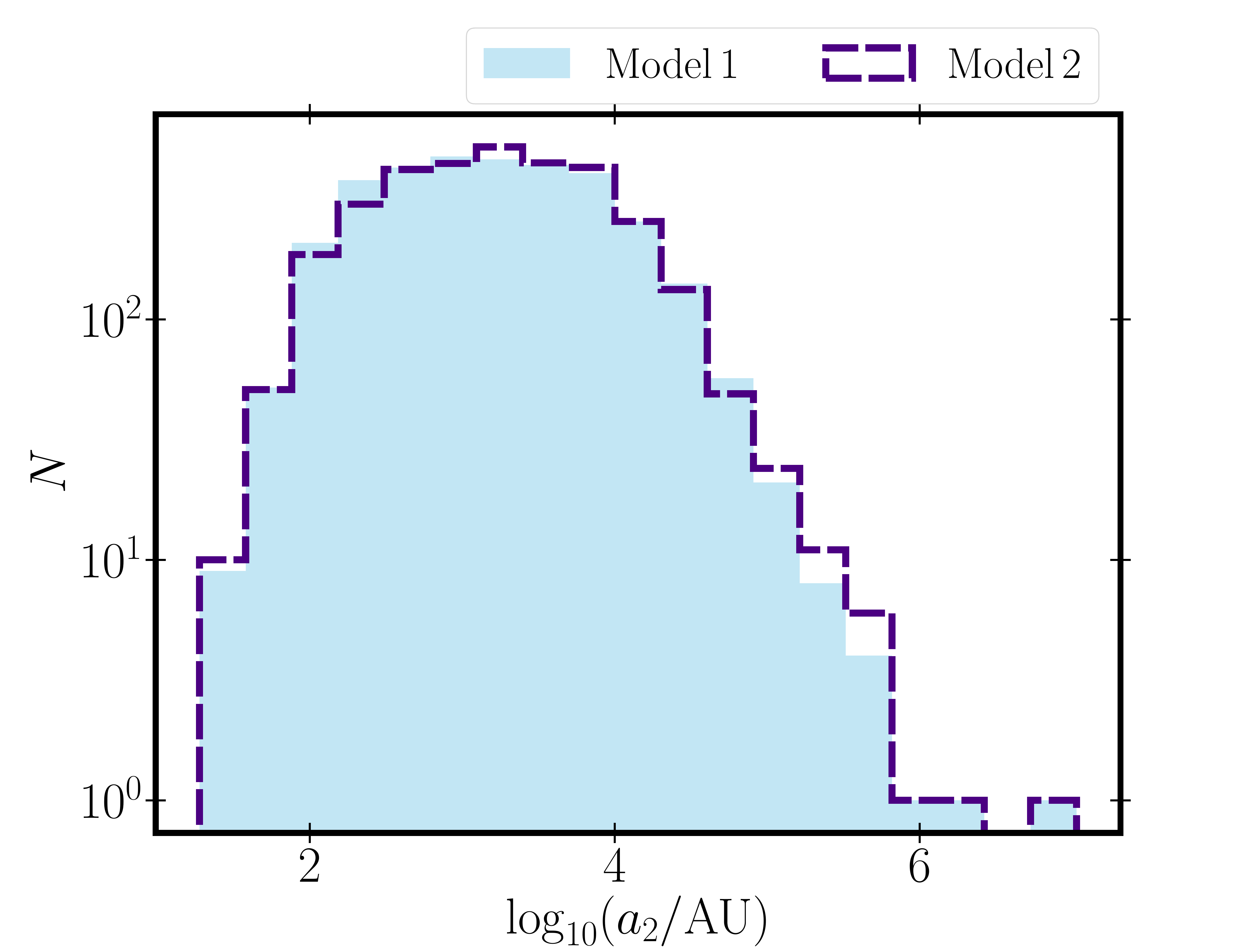}
 \caption{Outer semimajor axis distribution for systems in which the inner binary explodes as SNe Ia.}
 \label{fig:img13}
\end{figure}

In the isolated binary evolution channels, most SNe Ia explosions result from the formation of close WD binaries following CE evolution. However, in triple-star systems, in addition to CE evolution, the tertiary star can also aid the formation of SNe Ia.  Fig.\,\ref{fig:img9} represents the initial semimajor axis distribution of systems that explode as SNe Ia. From the semimajor axis distribution, it is evident that that the systems with wide semimajor axes undergo collisions triggered by high eccentricity oscillations in the inner binary due to the tertiary star. This is because the triples with wide inner binaries have shorter secular time-scales (all else being the same), whereas the short-range precession time-scales in the inner binaries are longer. Both these effects contribute to a larger probability for exciting high eccentricities in the inner binary. The ZLK mechanism produces high amplitude eccentric oscillations in systems with high intial mutual inclinations. Furthermore, from Fig.\,\ref{fig:img10}, we can see that the systems with high initial mutual inclinations are more likely to undergo eccentric collisions than circular mergers. There is a strong decrease in the number of systems undergoing circular mergers for higher inclinations near 90 degrees (note that Fig. 10 uses a log scale). Such a strong dip is not apparent in the systems undergoing eccentric collisions. Whereas, in case of the opposide sides of the inclination distribution, inner orbits are more likely brought closer together by CE evolution, thereby inducing circular mergers. Systems undergoing circular mergers are dominant; their contribution to the total SNe Ia is found to be $(99.0 \pm 1.7)\%$ and $(99 \pm 2)\%$ for $\mathrm{Model\,1}$ and $\mathrm{Model\,2}$ respectively. There are also systems undergoing eccentric collisions, but with a smaller contribution. Their fractional contribution is found to be $(0.8 \pm 0.1)\%$ for $\mathrm{Model\,1}$, and $(1.0 \pm 0.1)\%$ for $\mathrm{Model\,2}$.

\subsection{DD and SD SNe Ia\label{subsec:SD_DD}}

It is interesting to analyze the contribution of DD and SD channels to SNe Ia. From our study, the DD channel surpasses the SD channel in great numbers. The percentage of systems undergoing DD SNe Ia is $(99 \pm 2)\%$  and $(99 \pm 2)\%$ for $\mathrm{Model\,1}$ and $\mathrm{Model\,2}$ respectively whereas the SD SNe is $(0.7 \pm 0.1)\%$  for $\mathrm{Model\,1}$ and $(0.7 \pm 0.1)\%$ for $\mathrm{Model\,1}$. When we carried out an in-depth analysis into the various sub channels contributing to DD SNe Ia, we find that the majority of them are He WD and CO WD mergers, and there is also a non-negligible contribution from CO-CO WD mergers. Fig.\,\ref{fig:img11} shows the relative contributions from He-CO WDs and CO-CO WDs. The detail numbers are displayed in Table\,\ref{table:Table 4}. The contribution of SD sub channels is analyzed in the next section.

\subsection{Chandrasekhar and Sub-Chandrasekhar mass SNe Ia\label{subsec:SCMandCM}}

As described in Section \ref{sec:SNe_pres}, there are two possible scenarios of single degenerate SNe Ia taken into account in our simulations. The first scenario involves the accretor (CO WD) gaining mass by accreting mass from a non-degenerate donor star and exploding as SNe Ia when the WD reaches the Chandrashekar mass limit (1.44\,$\mathrm{M}_\odot$). In the second case, the CO WD undergoes stable accretion via Roche lobe overflow. Here, the donor is a H poor, He burning stripped star. The WD undergoes double detonation and explodes as SNe Ia, well before the Chandrasekhar mass is reached. These two scenarios involve different evolutionary pathways (see, e.g., \citealt{2020IAUS..357....1R}). From our $\mathrm{Model\,1}$ simulations of single degenerate SNe Ia, the percentage of systems undergoing Chandrashekhar mass and sub-Chandrashekhar mass are $(9 \pm 5)\%$ and $(90 \pm 20)\%$ respectively. From the $\mathrm{Model\,2}$ simulations, the percentage of systems undergoing Chandrashekhar mass and sub-Chandrashekhar mass are $(14 \pm 6)\%$  and $(86 \pm 19)\%$ respectively. It is apparent from the statistics that the contribution of sub-Chandrasekhar mass SNe Ia dominate the fraction of single degenerate SNe Ia over Chandrasekhar mass SNe Ia.

\begin{table*}
\begin{tabular}{c c c c c c c c} 
\toprule
Models & Channel  & DD\tnote{1}  & SD\tnote{2}  & SCM\tnote{3}  & CM\tnote{4} & He+CO\tnote{5} &CO+CO\tnote{6}\\
 &  &  $(\%)$ & $(\%)$  & $(\%)$ & $(\%)$ & $(\%)$  & $(\%)$\\
\midrule
\multirow{2}{*}{Model\,1}&Triple  & $99.3 \pm 1.7$ & $0.7 \pm 0.1$ & $90.7 \pm 20.06$ & $9.3 \pm 4.9$ & $96.6 \pm 1.7$ & $3.4 \pm 0.2$\\
&Binary  & $99.4 \pm 1.9$ & $0.6 \pm 0.1$ & $93.7 \pm 23.8$ & $6.3 \pm 4.5$ & $99.3 \pm 1.9$ & $0.7 \pm 0.1$\\
\midrule
\multirow{2}{*}{Model\,2}&Triple&  $99.1 \pm 1.7$ & $0.9 \pm 0.1$ & $86.4 \pm 19.1$ & $13.6 \pm 5.9$ & $96.7 \pm 1.7$ & $3.3 \pm 0.2$\\
&Binary  & $99.3 \pm 1.9$ & $0.7 \pm 0.1$ & $90.0 \pm 20.7$ & $10.0 \pm 5.2$ & $99.3 \pm 1.9$ & $0.7 \pm 0.1$\\
\midrule
{Model\,3}&Triple  & $99.5 \pm 4.3$ & $0.5 \pm 0.2$ & $75.0 \pm 57.3$ & $25.0 \pm 28.0$ & $97.7 \pm 4.2$ & $2.3 \pm 0.5$\\
{Model\,4}&Triple & $100.0 \pm 6.8$ & $0.0$ & $0.0$ & $0.0$ & $79.7 \pm 5.8$ & $20.3 \pm 2.4$\\
{Model\,5}&Triple & $99.7 \pm 3.4$ & $0.3 \pm 0.1$ & $80.0 \pm 53.7$ & $20.0 \pm 21.9$ & $97.4 \pm 3.4$ & $2.6 \pm 0.4$\\
{Model\,6}&Triple & $99.5 \pm 3.4$ & $0.5 \pm 0.1$ & $66.7 \pm 43.03$ & $33.3 \pm 27.2$ & $96.8 \pm 3.4$ & $3.2 \pm 0.4$\\
\bottomrule
\end{tabular}

\begin{tablenotes}\footnotesize

\item DD - Percentage of DD SNe Ia of the total number of SNe Ia 
\item SD - Percentage of SD SNe Ia of  the total number of SNe Ia
\item SCM - Percentage of sub-Chandrasekhar mass SD SNe Ia of the total number of SD SNe Ia
\item CM - Percentage of Chandrasekhar mass SD SNe Ia of  the total number of SD SNe Ia
\item He+CO - Percentage of mergers/collisions of He WD and CO WD of the total number of DD SNe Ia
\item CO+CO - Percentage of mergers/collisions of CO WD and CO WD of the total number of DD SNe Ia

\end{tablenotes}

\caption{Contribution of different progenitors to SNe Ia. The `Binary' channel here refers to the inner binaries of the triple population, evolved without the tertiary star.}
\label{table:Table 4}
\end{table*}

\subsection{Isolated binary evolution\label{subsec:IBE}}

In order to understand the contribution of the triple evolution channel to the SNe Ia rate, it is important to compare to the rate attributed to the binary evolution channel. We constructed our initial binary population to be fully consistent with our sampling of the triple population. Specifically, we assumed the \citet{2001MNRAS.322..231K} IMF for the primary mass distribution; the secondary mass, eccentricity, and period distributions follow the functional distributions from \citet{2017ApJS..230...15M}. We used the same SNe Ia prescriptions as for our triple runs and, for consistency, the same population synthesis code MSE to evolve these binary systems for 10 Gyr. The estimated time integrated SNe Ia rate from our binary system calculations is ($3.2 \pm 0.1) \times {10^{-4}}$ $\mathrm{M}_\odot^{-1}$. The corresponding DTD from the binary evolution channel is shown in Fig.\,\ref{fig:img7} with the red dotted line.

\subsection{Properties of the third star\label{subsec:thirdstar}}

We examined the nature of the third star at the moment when the inner binary explodes as SNe Ia. We find that, in about 1 per cent of systems, two unbound stars collide during a phase of dynamical instability leading to SNe Ia, 20 per cent of stars are newly formed binaries as result of the merger of the inner binary and the tertiary component, 30 per cent of systems have an unbound third star, and 49 per cent of systems have a bound third star. The mass distribution of the tertiary star in those cases when it is still bound at the moment of the SNe Ia explosion is shown in Fig.\,\ref{fig:img12}; most of the tertiaries have masses less than 1 $\mathrm{M}_\odot$, peaking at $\approx 0.5 \, \mathrm{M}_\odot$. The outer semimajor axis distribution for the cases when the tertiary star is still bound at the time of the SNe Ia explosion is shown in Fig.\,\ref{fig:img13}; the outer semimajor axis is found to be distributed broadly up to more than a million au, but peaking at a semimajor axis of about $10^3$ au. In most of the cases, fly-bys are responsible for these wide orbits while in a small number of cases, CE evolution in the inner binary can also create these wide orbits. It is also noted that these outer orbits are mostly eccentric and expected to be typically short-lived. In addition, we point out that it would be very difficult to detect these ultra-wide orbits.

\section{Discussion} \label{sec:Discussions}

\begin{figure}
 \includegraphics[width=1\columnwidth]{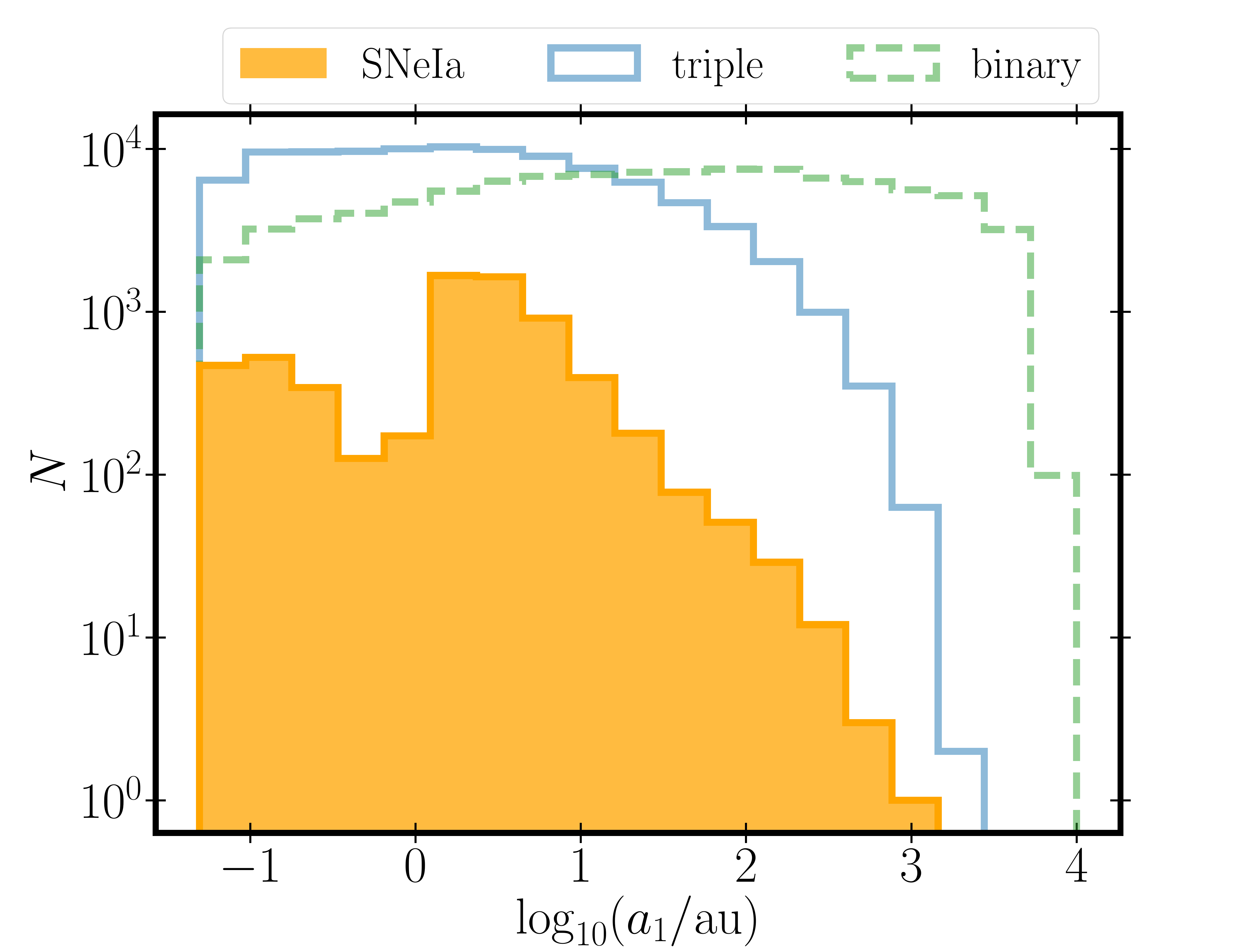}
 \caption{Initial inner semimajor axis distribution of our triple population (Model 1 - solid blue line), isolated binary population (dashed green line) and systems that explode as SNe Ia from triple evolution channels (filled yellow columns).}
 \label{fig:img14}
\end{figure}

\subsection{The effect of the tertiary star\label{subsec:Effectofthirdstar}}

In order to investigate the effect of the tertiary on the formation of SNe Ia, we investigated three different data sets: runs with the hierarchical triple population, runs with the inner binary of the triples population (after removing the tertiary star from the triple system) and runs with the isolated binary population. We find that the time-integrated SNe Ia rates from hierarchical the triple population, inner binary of triple population and isolated binary population cases are $(3.60 \pm 0.04) \times {10^{-4}} \mathrm{M_{\odot}^{-1}}$, $(2.90 \pm 0.04) \times {10^{-4}} \mathrm{M_{\odot}^{-1}}$, and $(3.2 \pm 0.1) \times {10^{-4}} \mathrm{M_{\odot}^{-1}}$, respectively. This shows that the hierarchical triple population slightly yields the highest contribution to the SNe Ia rate. 

The tertiary star is contributing to the SNe Ia in different ways: Firstly, the stability configuration of the hierarchical triple population demands the inner binaries of the hierarchical triple population to be in tighter orbits than those of the isolated binary population. We can also see from Fig.\,\ref{fig:img14} that these tight inner binaries contribute the most to SNe Ia explosions. Secondly, the tertiary star can assist in shrinking the inner binary orbit and bringing them closer to lead to either a circular merger via tides, stable mass transfer,  and/or CE, or an eccentric collision. One can see these two contribution of the tertiary in two different peaks in the yellow column of Fig.\,\ref{fig:img14}.

\citet{2013MNRAS.430.2262H} carried out a similar study by restricting the initial conditions to systems with ${{a_1}{(1-e_1^2)}} > {12 \mathrm{\,au}}$ and estimated the SNe Ia rates from triples to be on the order of $10^{-6}\,\mathrm{M_{\odot}^{-1}}$. Our rates agree with \citet{2013MNRAS.430.2262H} for systems with ${{a_1}{(1-e_1^2)}} > {12 \mathrm{\,au}}$. The isolated binary population rates from our calculations are similar to that of \citet{2014A&A...563A..83C}. Putting our results in perspective, the time integrated rates from the isolated binary and triple channels in our simulations are $(3.2 \pm 0.1) \times {10^{-4}} \mathrm{M_{\odot}^{-1}}$  and $(3.60 \pm 0.04) \times {10^{-4}} \mathrm{M_{\odot}^{-1}}$ respectively. The observed time integrated rate from \citet{2012MNRAS.426.3282M} is $(1.3 \pm 0.2) \times {10^{-3}} \mathrm{M_{\odot}^{-1}}$. The combined rates from the triple and binary channels thus contribute to about $\sim52$ per cent of the observed rate, of which the largest contribution comes, somewhat surprisingly, from triple systems. The discrepancy between the observed and theoretical rates demands the exploration of other SNe Ia progenitors, though it should be noted that there are significant uncertainties in our models, which we address in the next Section.

\subsection{Uncertainties in the models\label{subsec:uncertainties}}

\begin{figure}
 \includegraphics[width=1\columnwidth]{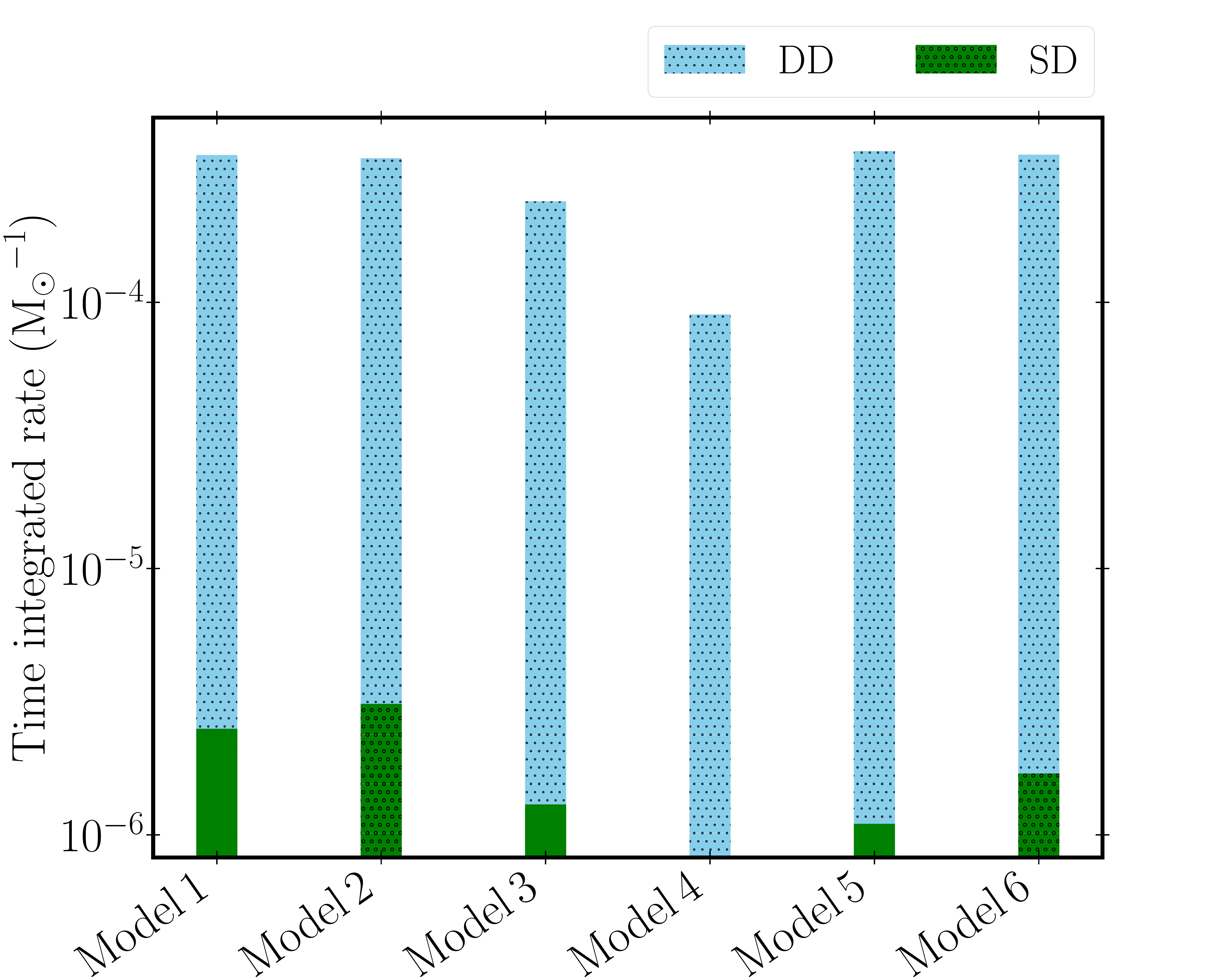}
 \caption{Impact of different models on the SNe Ia rate. See Table\,\ref{table:Table 1} for a description of the models.}
 \label{fig:img15}
\end{figure}

Fig.\,\ref{fig:img7} shows the DTD of DD SNe Ia with solid blue line, dashed green line, and dotted red line representing the corresponding DTD from Model 1, Model 2, and isolated binary population, respectively. From Fig.\,\ref{fig:img7}, it is evident that the total rate and DTD are not affected by the underlying $q_\mathrm{out}$ distribution. However, as shown by Fig.\,\ref{fig:img15}, the CE efficiency parameter $\alpha_\mathrm{CE}$ does strongly affect the number of SNe Ia and hence the rates. A higher efficiency parameter ($\alpha_\mathrm{CE}$ = 10) results in more early mergers and thereby fewer SNe Ia than a lower efficiency ($\alpha_\mathrm{CE}$ = 1). Furthermore, a low efficiency parameter ($\alpha_\mathrm{CE}$ = 0.1) results in less transfer of orbital energy during CE evolution, thereby reducing the number of close binaries that could lead to SNe Ia explosions. The effects of fly-bys and lower metallicity are negligible in producing SNe Ia. We have used multiplicity fractions from \citet{2017ApJS..230...15M} while normalizing the rates for every models. Uncertainties in the multiplicity fractions, as well as the precise values for the upper and lower mass of the synthesized population, propagate into errors in the mass normalisation, which we have not considered here for simplicity.  We further note as a caveat that all uncertainties that apply to single and binary star evolution (including the criteria for what exactly produces a SNe Ia transient), also apply here.

\subsection{Predominance of circular mergers\label{subsec:circularmergers}}

As described in Section 4.2, according to our results, SNe Ia attributed to circular mergers via CE are dominant compared to those following eccentric collisions. This is in strong contrast to \citet{2015MNRAS.454L..61D}, who suggested that head-on WD collisions in isolated triples are the dominant channel for producing SNe Ia, and agrees with previous works \citep{2013MNRAS.430.2262H,2018A&A...610A..22T} which found that the rate of head-on collisions in WD-WD systems is too low to explain the observed SNe Ia rate.

\begin{figure}
 \includegraphics[width=1\columnwidth]{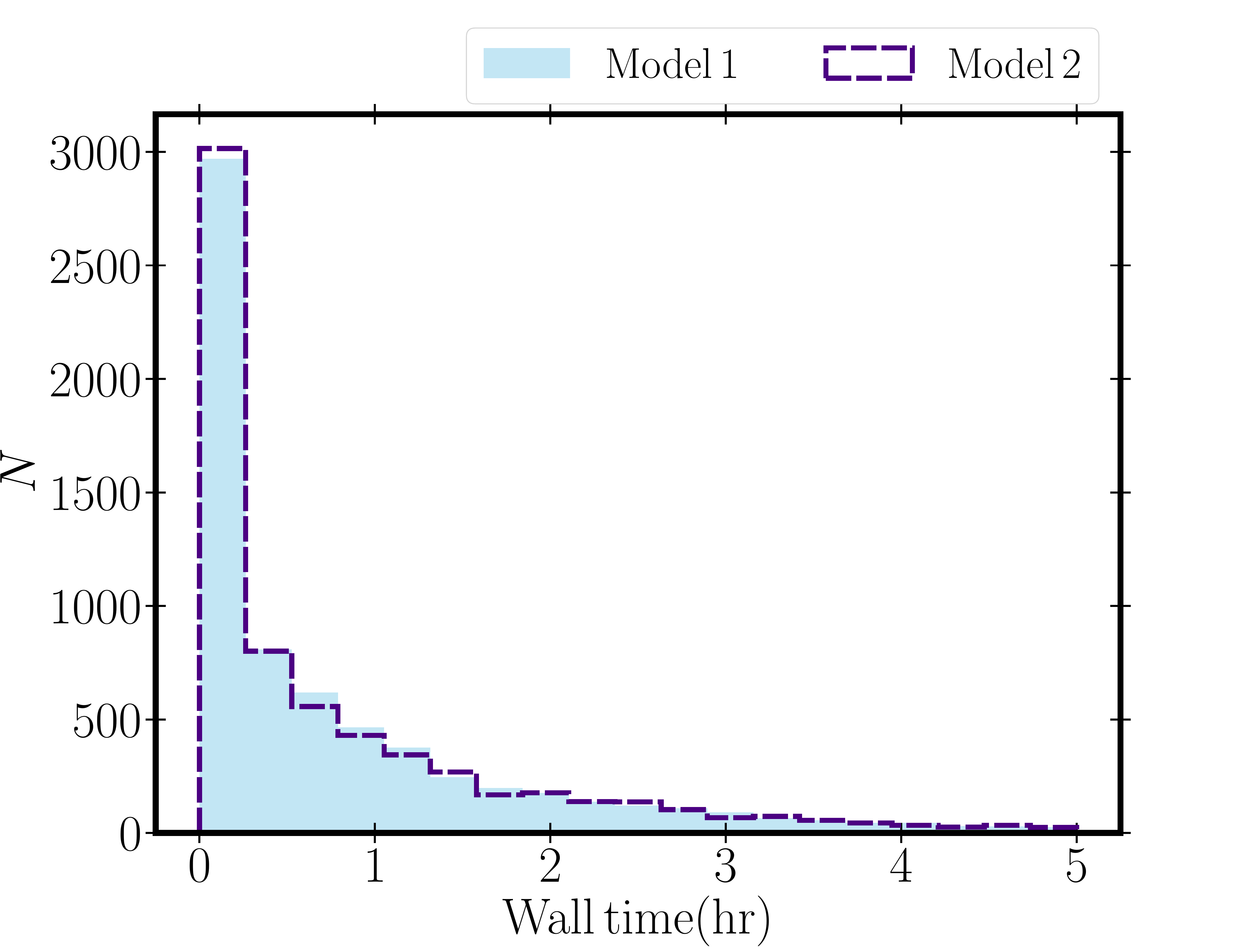}
 \caption{Wall time for systems that explode as SNe Ia.}
 \label{fig:img16}
\end{figure}

About $6\%$ of total evolved triple systems did not complete due to the set restricted wall time of 5 hrs. Fig.\,\ref{fig:img16} shows the wall time distribution of systems that explode as SNe Ia. It is evident that the majority of the systems that explode as SNe Ia have a wall time within 1 hr; any contribution from systems with longer wall-time is negligible.

Even though the SNe Ia rate from the triple evolution channel is found to contribute similarly to that of binary evolution channel, the combined rate from triples and binaries is still inadequate to explain the complete the observed rate. Thus, a detailed study on other possible SNe Ia progenitors and contribution of SNe Ia from higher order systems should be done in the future.

\section{Conclusions} \label{sec:conclusions}

We studied rates of SNe Ia explosions in hierarchical triple systems by performing evolutionary population synthesis calculations. The triple populations were constructed following \citet{2001MNRAS.322..231K} and \citet{2017ApJS..230...15M}, and we only considered initially dynamically stable systems, using the stability criterion from \citet{2001MNRAS.321..398M}. The systems in which one or more of the stars are filling their Roche lobe at the beginning of the main sequence were ignored. Our sampled triples were evolved using the evolutionary population synthesis code MSE \citep{2021MNRAS.502.4479H} for a period of 10 Gyr, and a statistical analysis was carried out. The results are summarized as follows.

\begin{enumerate}
  \item We found 5 unique formation channels to produce SNe Ia.
    \begin{enumerate}
        \item Unbound tertiary: a triple evolution channel in which the tertiary gets unbound when it collapses into a neutron star. The other reasons for the tertiary star to get unbound includes dynamical instability, and CE in the inner binary system.
        \item Double merger: a triple system in which the inner binary components merge to form a new massive star, which then later interacts with the tertiary star to produce a SNe Ia explosion.
        \item TCE: a triple system in which the massive tertiary transfers mass on top of the inner binary resulting in exchange, merger or dynamical instability. The newly formed components then interact to produce a SNe Ia explosion at a later time.
        \item Eccentric collision: in a triple star system, when the mutual inclinations are large, the inner binary undergoes eccentricity enhancements due to secular evolution. This can increase the eccentricity of the inner binary, of which the components then collide to form a SNe Ia.
        \item Dynamical instability: unlike the isolated binary channel, this triple evolution channel can produce SNe Ia without undergoing a CE phase. This is a purely dynamical channel in which the the triple system undergoes dynamical instability to explode as SNe Ia.
    \end{enumerate}
  \item Head-on eccentric collisions of WDs contribute only about 1 per cent (Model 1) to the total SNe Ia, while the rest are all circular mergers that involve CE evolution.
  \item When there occurs a SNe Ia in the inner binary, the third star is found to be bound in $\sim49$ per cent of systems. The mass distribution of the bound star peaks around 0.5 $\mathrm{M}_\odot$, and the outer orbital semimajor axis is distributed over a broad range of about few million au, peaking at about $10^3\,\mathrm{au}$.
  \item We estimated the delay time distribution for single and DD SNe Ia, which was presented in Fig.\,\ref{fig:img7}, and Fig.\,\ref{fig:img8} respectively. The time-integrated rate of SNe Ia from the triple evolution channel is found to be slightly higher than that of the binary evolution channel, although this conclusion is affected by uncertainties in the models.
  \item Previously, \citep{2013MNRAS.430.2262H,2018A&A...610A..22T} considered triples with only wide inner binaries and found a comparatively low contribution of triples to the SNe Ia rate. However, when the complete set of parameters is included, it is evident that the triple channel is an important channel in producing SNe Ia explosions.
  \item According to our models, the combined rate from the triple and binary evolution channels contributes to about $52$ per cent of the observed SNe Ia rate.
\end{enumerate}

\section{Acknowledgements} \label{sec:Acknowledgements}
ASH thanks the Max Planck Society for support through a Max Planck Research Group. ASR thanks Or Graur for his inputs on observational references.

\appendix

\section{Mobile diagrams\label{appendix:mobilediagrams}}
We use mobile diagrams to describe the different evolutionary stages of triple star evolution in MSE. The blue and orange boxes represent the inner and outer orbits of the triple system, respectively. Colors indicate the evolutionary stage following the stellar types defined in \citet{2000MNRAS.315..543H}. The legend explains the colors of different stellar types, and the acronyms are described in  Table.\,\ref{table:Table 5}. The red arrows pointing from one star to the other represent strong interactions such as stable mass transfer, unstable mass transfer, and collisions. The orange and red shaded regions around stars represent Roche lobe overflow and CE episodes, respectively. The star symbol in the final panel of every mobile diagram shows a SNe Ia explosion. Every SNe Ia explosion involves only two stars.

\setcounter{table}{0}
\begin{table*}
\caption{Mobile diagram Acronyms}
\begin{tabular}{c c c c c c c c c c c c c c c c c } 
\toprule
Acronym & Stellar type \\
\midrule
low-mass MS & Main Sequence star ($M$ $\lesssim$ 0.7 $\mathrm{M}_\odot$) \\
MS & Main Sequence star ($M$ $\gtrsim$ 0.7 $\mathrm{M}_\odot$) \\
HG & Hertzsprung Gap  \\
RGB & Red giant branch \\
CHeB & Core He burning \\
EAGB & Early Asymptotic giant branch \\
TPAGB & Thermally pulsating asymptotic giant branch\\
HeMS & He Main Sequence \\
HeHG & Helium Hertzsprung Gap\\
HeGB & Helium Giant Branch \\
HeWD & He WD\\
COWD & CO WD\\
ONeWD& ONe WD\\
NS& Neutron star\\
BH& Black hole\\
\bottomrule
\end{tabular}
\label{table:Table 5}
\end{table*}

\begin{table*}
\caption{Contribution of different progenitors to SNe Ia. The `Binary' channel here refers to the inner binaries of the triple population, evolved without the tertiary star.}
\begin{tabular}{c c c c c c c c} 
\toprule
Models & Channel  & DD\tnote{1}  & SD\tnote{2}  & SCM\tnote{3}  & CM\tnote{4} & He+CO\tnote{5} &CO+CO\tnote{6}\\
&  &  ($10^{-4}\mathrm{M_\odot}^{-1}$) & ($10^{-4}\mathrm{M_\odot}^{-1}$)  & ($10^{-4}\mathrm{M_\odot}^{-1}$) & ($10^{-4}\mathrm{M_\odot}^{-1}$) & ($10^{-4}\mathrm{M_\odot}^{-1}$)  & ($10^{-4}\mathrm{M_\odot}^{-1}$)\\
\midrule
\multirow{2}{*}{Model\,1}&Triple  & $3.57 \pm 0.04$ & $0.025 \pm 0.003$ & $0.024 \pm 0.004$ & $0.0009 \pm 0.0007$ & $3.45 \pm 0.04$ & $0.122 \pm 0.008$\\
&Binary  & $2.88 \pm 0.04$ & $0.017 \pm 0.003$ & $0.017 \pm 0.003$ & $0.0001 \pm 0.0002$ & $2.86 \pm 0.04$ & $0.020 \pm 0.003$\\
\midrule
\multirow{2}{*}{Model\,2}&Triple&  $3.47 \pm 0.04$ & $0.031 \pm 0.004$ & $0.030 \pm 0.004$ & $0.0010 \pm 0.0007$ & $3.35 \pm 0.04$ & $0.114 \pm 0.008$\\
&Binary  & $2.88 \pm 0.04$ & $0.020 \pm 0.003$ & $0.020 \pm 0.003$ & $0.0001 \pm 0.0003$ & $2.86 \pm 0.04$ & $0.020 \pm 0.003$\\
\midrule
{Model\,3}&Triple  & $2.39 \pm 0.07$ & $0.013 \pm 0.005$ & $0.013 \pm 0.005$ & $0.0003 \pm 0.0008$ & $2.33 \pm 0.071$ & $0.054 \pm 0.011$\\
{Model\,4}&Triple & $0.90 \pm 0.04$ & $0.0$ & $0.0$ & $0.0$ & $0.72 \pm 0.04$ & $0.183 \pm 0.020$\\
{Model\,5}&Triple & $3.69 \pm 0.09$ & $0.011 \pm 0.005$ & $0.010 \pm 0.005$ & $0.0003 \pm 0.0008$ & $3.59 \pm 0.09$ & $0.097 \pm 0.014$\\
{Model\,6}&Triple & $3.58 \pm 0.09$ & $0.017 \pm 0.006$ & $0.017 \pm 0.006$ & $0.0006 \pm 0.0011$ & $3.47 \pm 0.09$ & $0.114 \pm 0.016$\\
\bottomrule
\end{tabular}

\begin{tablenotes}\footnotesize

\item DD - Rate of DD SNe Ia of the total number of SNe Ia 
\item SD - Rate of SD SNe Ia of  the total number of SNe Ia
\item SCM - Rate of sub-Chandrasekhar mass single degenerate SNe Ia 
\item CM - Rate of Chandrasekhar mass single degenerate SNe Ia 
\item He+CO - Rate of mergers/collisions of He WD and CO WD 
\item CO+CO - Rate of mergers/collisions of CO WD and CO WD 

\end{tablenotes}

\label{table:Table 4}
\end{table*}

\clearpage
\bibliography{sample631}{}
\bibliographystyle{aasjournal}



\end{document}